\documentclass[showpacs,preprintnumbers,amsmath,amssymb,floatfix]{revtex4}

\usepackage{graphicx}
\usepackage{dcolumn}
\usepackage{bm}

\begin{document}

\def\bbox#1{\hbox{\boldmath${#1}$}}
\def\gtsim{$\raisebox{0.6ex}{$>$}\!\!\!\!\!\raisebox{-0.6ex}{$\sim$}\,\,$}
\def\ltsim{$\raisebox{0.6ex}{$<$}\!\!\!\!\!\raisebox{-0.6ex}{$\sim$}\,\,$}
\def\pt{\bbox{p}_t}
\def\xx{\hbox{\boldmath{$ x $}}}
\def\pp{\hbox{\boldmath{$ p $}}}
\def\pt{\pp_{{}_T}}
\def\yb{{\bar y }}
\def\mt{m_{{}_T}}
\def\zb{{\bar z }}
\def\tb{{\bar t }}
\def\rhoe{ {\rho_{\rm eff}} }
\def\qq{\hbox{\boldmath{$ q $}}}

\title{ Chaoticity Parameter $\lambda$ in Hanbury-Brown-Twiss
Interferometry}

\author{Cheuk-Yin Wong}

\affiliation{Physics Division, Oak Ridge National Laboratory, 
Oak Ridge,\footnote{Electronic address: wongc@ornl.gov}TN  37831 USA}

\author{Wei-Ning Zhang }

\affiliation{ School of Physics \& Optoelectronic Technology,Dalian
University of Technology,Dalian,\footnote{Electronic address:
wnzhang@dlut.edu.cn}116024 China\\ Physics Department, Harbin
Institute of Technology, Harbin 150006, China}

\received{\today}

\begin{abstract}
In Hanbury-Brown-Twiss interferometry measurements using identical
bosons, the chaoticity parameter $\lambda$ has been introduced
phenomenologically to represent the momentum correlation function at
zero relative momentum.  It is useful to study an exactly solvable
problem in which the $\lambda$ parameter and its dependence on the
coherence properties of the boson system can be worked out in great
detail.  We are therefore motivated to study the state of a gas of
noninteracting identical bosons at various temperatures held together
in a harmonic oscillator potential that arises either externally or
from bosons' own mean fields.  We determine the degree of
Bose-Einstein condensation and its momentum correlation function as a
function of the attributes of the boson environment.  The parameter
$\lambda$ can then be evaluated from the momentum correlation
function.  We find that the $\lambda(p,T)$ parameter is a sensitive
function of both the average pair momentum $p$ and the temperature
$T$, and the occurrence of $\lambda=1$ is not a consistent measure of
the absence of a coherent condensate fraction.  In particular, for
large values of $p$, the $\lambda$ parameter attains the value of
unity even for significantly coherent systems with large condensate
fractions.  We find that if a pion system maintains a static
equilibrium within its mean field, and if it contains a
root-mean-squared radius, a pion number, and a temperature typical of
those in high-energy heavy-ion collisions, then it will contain a
large fraction of the Bose-Einstein pion condensate.

\end{abstract}


\maketitle

\section{Introduction}
In high-energy collision processes, Hanbury-Brown-Twiss (HBT)
intensity interferometry \cite{Hbt54} has been used to study the
space-time geometry of the source of particles
\cite{Won94}-\cite{Won07}.  As is well known, for identical bosons the
interference phenomenon arises from Bose-Einstein correlations and
depends sensitively on the degree of coherence of the boson system
\cite{Gla63}.  The HBT correlation occurs for a chaotic source but not
for a coherent source.

In phenomenological measurements, one represents the correlation in
terms of the momentum correlation function
$C(\bbox{p}+\bbox{q}/2,\bbox{p}-\bbox{q}/2)=C(\bbox{p}_1,\bbox{p}_2)
=G^{(2)}(\bbox{p}_1,\bbox{p}_2;\bbox{p}_1,\bbox{p}_2)
/G^{(1)}(\bbox{p}_1,\bbox{p}_1)G^{(1)}(\bbox{p}_2,\bbox{p}_2)$, where
$\bbox{p}_1$ and $\bbox{p}_2$ are the momenta of the pion pair and
$G^{(i)}$ is the $i$-th-order pion density matrix.  One introduces the
parameter $\lambda =[C(q=0)-1]$ that is purported to represent the
degree of chaoticity of the pion medium and bears the name ``the
chaoticity parameter''.  Experimental measurements with pions
persistently indicate that the chaoticity parameter $\lambda$ is
substantially less than the value of unity for a fully chaotic source.
Some part of this reduction of the chaoticity parameter $\lambda$
from unity may be attributed to the occurrence of the decays of
long-live resonances \cite{Hei96}.  However as emphasized not the
least by Glauber \cite{Gla06}, part of the reduction of the
`chaoticity parameter' $\lambda$ from unity may arise from the
coherence of the pion gas.

Even though the chaoticity parameter has been widely used in all HBT
measurements in high-energy collisions, how the chaoticity parameter
can be determined theoretically has not been resolved.  The difficulty
is further compounded for heavy-ion collisions because the dynamics of
pions after their production in high-energy heavy-ion collisions is
very complicated.  The process of initial hadronization and the
subsequent interactions between pions are beyond the realm of
present-day knowledge.

It is therefore useful at this stage to study an exactly solvable
problem for which the $\lambda$ parameter can be determined explicitly
and the transition from the coherent phase to the chaotic phase can be
worked out in detail.  We are motivated to investigate the state of a
gas of noninteracting identical bosons held together in a harmonic
oscillator potential at various temperatures.  We shall study the
occurrence of Bose-Einstein condensation and the two-body momentum
correlation function as a function of the attributes of the boson gas in
such an environment.  This will allow us to examine explicitly the
transition from the coherent phase to the chaotic phase and to study
how this phase transition may affect the HBT measurements and the
$\lambda$ parameter for a set of known attributes of the Bose-Einstein
gas assembly.

In atomic physics, the harmonic oscillator potential introduced here
can arise from an external trap.  In high-energy heavy-ion collisions,
the harmonic oscillator potential can arise approximately from the
mean-field potential experienced by a pion, owing to the interactions
generated by other pions and medium particles.  Although the strength
of the pion mean field is not known at present, the results obtained
here will serve as useful supplementary tools to study the
circumstances in which a pion system may form a Bose-Einstein
condensate in heavy-ion collisions.  They will stimulate future
investigations on the magnitude of the pion mean-field potential and
pave the way for future investigations on momentum correlations for
pions under more complicated dynamical evolutions.

Pions produced in high-energy heavy-ion collisions have a temperature
that is of the order of the pion rest mass.  The motion of the pions
is relativistic and a proper treatment will need to be relativistic
in nature.  We shall carry out both a nonrelativistic and a
relativistic treatment of the pions  to understand what
features of the coherence are sensitively affected by the
relativistic motion.

Important advances in our understanding of the coherence properties of
identical bosons have been made recently in another related field, the
physics of atomic boson systems at low temperatures
\cite{Pol95,Nar99,Gom06,Yas96,Hel03,Gre05,Fol05,Ott05,Sch05,Est06}.
Theoretical and experimental work in atomic physics has focused on the
correlation function in the configuration space.  In particular, the
second-order correlation function $g^{(2)}(\bbox{r_1},\bbox{r_2})$ has
been obtained to give the probability of detecting a boson at
${\bbox{r}_1}$ in coincidence with the detection of another identical
boson at $\bbox{r}_2$ \cite{Nar99}.  From the shape of this
correlation function $g^{(2)}(\bbox{r}_1,\bbox{r}_2)$ as a function of
the relative separation $\bbox{r}_1-\bbox{r}_2$, the theoretical
spatial correlation length can be extracted.  Experimentally, the
measurements of various arrival times and positions at the detectors
in HBT interferometry are then used to determine the spatial
correlation length, for comparison with theoretical analyses
\cite{Sch05}.  We wish to adopt a treatment complementary to that in
atomic physics by examining the correlation function in momentum
space, the standard arena for Bose-Einstein correlation analysis in
high-energy nuclear collisions \cite{Won94}-\cite{Won07}.  Our
investigation of the correlation function in momentum space is greatly
facilitated by utilizing the results of the correlation function in
configuration space obtained in atomic physics
\cite{Pol95,Nar99,Gom06}.

With regard to the low-temperature measurements with atoms, the
perspectives of studying the correlation in momentum space presented
here offer useful complementary points of view. In momentum space
the trapped atoms are now described as having an equilibrium momentum
distribution, appropriate for the system in a given external field at
a given temperature.  The sudden removal the external field allows the
initial momentum distribution of the particle to be frozen at the
moment of the external field removal, as appropriate under the
application of the sudden approximation in quantum
mechanics. Subsequent free streaming of the particles without the
external field and mutual interactions allows the reconstruction of
the momentum distribution of the source at the moment of its freezing
out.  In atomic physics, the correlation function in the
complementary momentum space has many rich features as it is sensitive
to many kinematic variables and the geometry of the source particles.

This paper is organized as follows.  In Section II, we review the
degree of Bose-Einstein condensation as a function of temperature and
particle number.  In Section III, we study the one-body and two-body
momentum density matrices.  In Section IV, we express the momentum
correlation function in terms of the one-body momentum density matrix
and the ground-state wave function and also express the momentum
correlation function in terms of the Wigner function.  In Section V,
we evaluate the one-body density matrix and the Wigner function for
bosons in the harmonic oscillator potential. In Section VI, we study
the spatial and momentum distributions of these boson assemblies.  In
Section VII, we evaluate the momentum correlation function
$C(\bbox{p},\bbox{q})$, for different values of the average pair
momentum $\bbox{p}$ and temperature $T$. We extract the $\lambda$
parameter and the HBT radii.  In Section VIII we study the condensate
fraction for a nonrelativistic pion gas with a given root-mean-squared
radius in a static equilibrium at various temperatures.  In Section
IX, we investigate the relativistic treatment of the boson in a
harmonic oscillator potential.  In Section X, we evaluate the boson
spatial density and estimate its condensate fraction for the
relativistic boson gas. In Section XI, we present our conclusions and
discussions.

\section{Condensate fraction as a function of temperature}

We shall first review the theoretical work on the coherence and
correlations of identical bosons in atomic physics
\cite{Pol95,Nar99,Gom06} so as to pave the way for our investigation
of the correlation function in momentum space.  We consider first a
nonrelativistic gas of identical bosons in a harmonic potential at
temperature $T$ with the potential specified by
\begin{eqnarray}
\label{VV}
V({\bf r})=\frac{1}{2} m \omega^2 r^2 = \frac {1}{2} \hbar \omega \left ( \frac {r}{a}\right)^2,
\end{eqnarray}
where $m$ is the rest mass of a boson and $\hbar \omega$ measures
the strength of the external potential.  We shall measure lengths in
units of the harmonic oscillator length parameter
$a=\sqrt{\hbar/m\omega}$, momenta in units of $\hbar/a$, and energies
in units of $\hbar \omega$.

The states in the harmonic oscillator potential are characterized by
energy levels $\epsilon_n=(n +\frac{3}{2}) \hbar \omega$ with the
associated degeneracy of $g_n=(n+1)(n+2)/2$.  Following Ref.\
\cite{Nar99}, it is convenient to use the recalibrated energy level
$\tilde \epsilon_n = n\hbar \omega$ measured relative to $3\hbar/2$.

As the temperature of the gas is lowered below the condensation
temperature $T_c$, condensation of the nonrelativistic massive boson
gas occurs.  As is well known, the fluctuation of the number of
particles in the condensate state, the $n=0$ state, depends
sensitively on the assumed statistical ensemble.  A grand canonical
ensemble will lead to a condensate ground state number fluctuation
that is as large as the number of particles in the condensate ground
state, $n=0$.  The grand canonical ensemble cannot be used to describe
the number of particles $N_0$ in the ground state condensate.  The
condensation can best be studied in a canonical ensemble for the case
with a fixed number of particles \cite{Pol95}. Comparison of the
results from the canonical and the grand canonical ensemble in Ref.\
\cite{Pol95} indicates however that even though the number of
particles in the ground $n=0$ state can only be described by the
canonical ensemble, the occupation number distribution of the $n > 0$
harmonic oscillator states can be appropriately described by the grand
canonical ensemble with only very small corrections.  The difference
in these two ensembles for the $n>0$ states becomes very small as the
number of particles increases.

Therefore, for a fixed number of particles $N$ at a given temperature
$T/\hbar \omega$, we shall follow Refs.\ \cite{Pol95} and \cite{Nar99}
to determine the condensate configuration by the following
requirements: (i) a fixed total number of particles $N$ in a canonical
ensemble for the condensate $n=0$ state, and (ii) an occupation number
distribution in a grand canonical ensemble for the $n> 0$
states. Accordingly, we have the following three conditions to
determine the condensate configuration of the system with a fixed
number of particles $N$ at a temperature $T=1/\beta$:
\begin{eqnarray}
\label{eq2}
N=N_0+N_T,
\end{eqnarray}
where $N_0$ is the number of condensate particles in the $n=0$ state,
\begin{eqnarray}
\label{N0}
N_0=\frac{z}{1-z},
\end{eqnarray}
$N_T$ is the number of ``chaotic'' particles in the $n> 0$ states,
\begin{eqnarray}
\label{NT}
N_T=\sum_{n>0}^\infty \frac {g_n z e^{- \beta { \tilde \epsilon}_n}} {1-z
                e^{- {\beta \tilde \epsilon}_n}},
\end{eqnarray}
and $z$ is the fugacity parameter. For the harmonic oscillator
potential, the summation for $N_T$ can be carried out analytically and
Eq.\ (\ref{NT}) can be simplified to 
\begin{eqnarray}
\label{eq5}
N_T=\sum_{k=1}^\infty z^k \frac 
{ e^{- k \beta \hbar \omega}(3-3e^{- k \beta \hbar \omega}+
                           e^{- 2k \beta \hbar \omega})}
{(1-e^{- k \beta \hbar \omega})^3}.
\end{eqnarray}
Equations (\ref{eq2})-(\ref{NT}) can be reduced into a single condensate
configuration condition,
\begin{eqnarray}
\label{eq6}
N=\frac{z}{1-z}+\sum_{k=1}^\infty z^k \frac 
{ e^{- k \beta \hbar \omega}(3-3e^{- k \beta \hbar \omega}+
                           e^{- 2k \beta \hbar \omega})}
{(1-e^{- k \beta \hbar \omega})^3}.
\end{eqnarray}

\begin{figure}[h]
\includegraphics[scale=0.5]{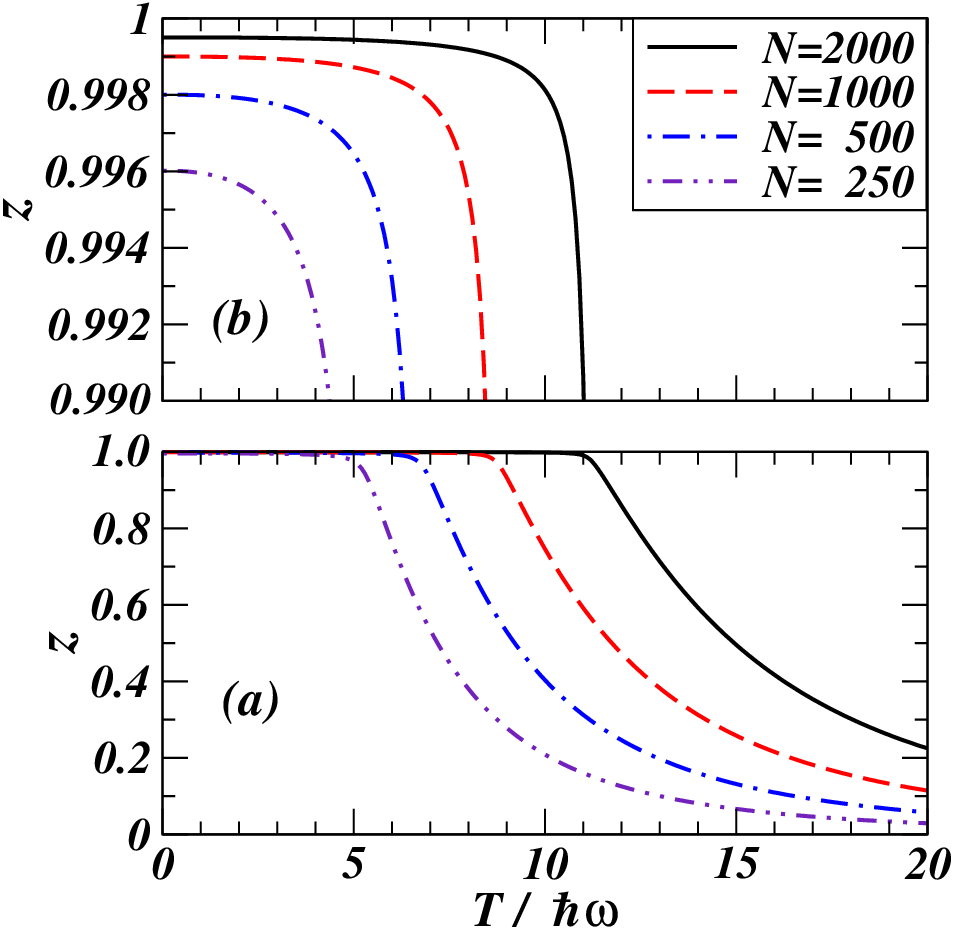}
\caption{(Color online) (a) The fugacity parameter $z$ satisfying the
condensate configuration condition Eq.\ (\ref{eq6}) for different
boson numbers $N$, as a function of temperature $T/\hbar \omega$
and ($b$) an expanded view in the $z\sim 1$
region.  }
\end{figure}

Because $N$ and $\beta \hbar \omega=\hbar\omega /T$ are fixed, the above
condensate configuration condition can be solved numerically to
determine the unknown $z$ (by Newton's method with fast convergence).
After the value of the solution $z$ is obtained, $N_0$ and $N_T$ can
be subsequently determined from Eqs.\ (\ref{N0}) and (\ref{eq5}) to
give the condensate configuration specified by the condensate fraction
$f_0$ and the `chaotic' fraction $f_T$,
\begin{eqnarray}
f_0=\frac {N_0}{N} {\rm ~~~and~~~~}f_T=\frac {N_T}{N}.
\end{eqnarray}
We show in Fig.\ 1 the fugacity solution $z$ which satisfies the
condensate configuration condition Eq.\ (\ref{eq6}) for different
temperatures $T/\hbar \omega$ and boson numbers $N$.  To get a better
view of the $z$ values, we show an expanded view of Fig.\ 1($a$) in
the $z\sim 1$ region in Fig. 1($b$).  We observe that the fugacity
parameter $z$ is close to unity in the strongly coherent region at low
temperatures.  In fact, the fugacity parameter $z$ at $T=0$ assumes
the value
\begin{eqnarray}
z(T=0)=\frac {N}{N+1}.
\end{eqnarray}
For a given boson number $N$, as the temperature increases from $T=0$,
the fugacity $z$ decreases very slowly in the form of a plateau until
the condensate temperature $T_c$ is reached, and it decreases very
rapidly thereafter.  The greater the number of bosons $N$, the greater
is the plateau region, as shown in Fig.\ 1($b$).  For example, for
$N=2000$ the value of $z$ is close to unity for $0<T/\hbar \omega<11$
in the plateau, and it deviates from unity substantially only for
temperatures $T/\hbar \omega >> 11$.

\begin{figure}[h]
\includegraphics[scale=0.5]{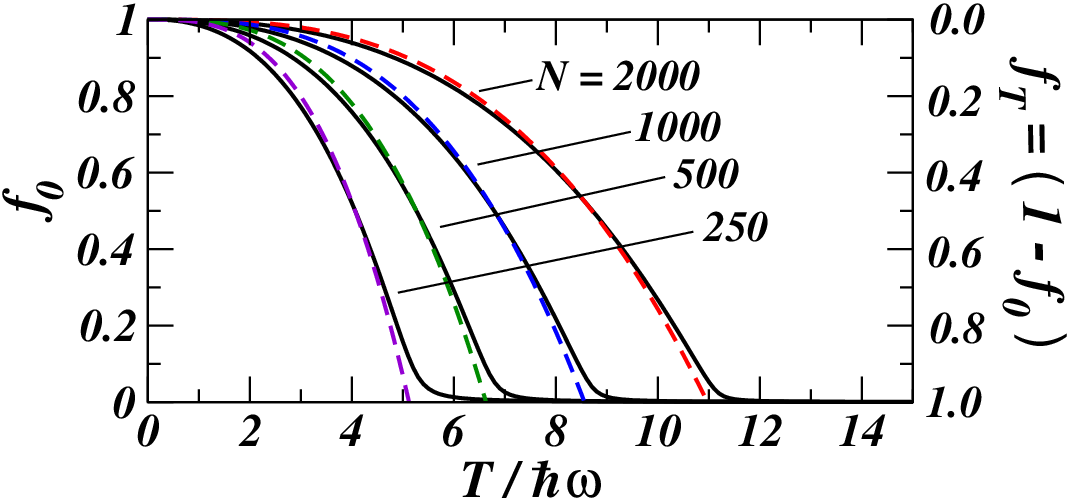}
\caption{(Color online) Solid curves represent the condensate fractions
$f_0(T)$, calculated with the condensate configuration condition Eq.\
(\ref{eq6}), as a function of $T/\hbar \omega$ for different boson
numbers $N$.  The abscissa labels for the corresponding chaotic
fraction $f_T(T)=[1-f_0(T)]$ are indicated on the right.  The dashed
curves are the fits to the solid curve results of $f_0(T)$ with the
function $1-(T/T_c)^3$ of Eq.\ (\ref{fit}) where the values of
$T_c/\hbar \omega$ for different $N$ values are listed in Table I.  }
\end{figure}

The condensate fractions $f_0(T)$ calculated with the fugacity
parameter of Fig.\ 1 for different boson numbers $N$ are represented
by solid curves in Fig.\ 2, as a function of $T/\hbar \omega$.  The
abscissa labels for the corresponding chaotic fraction
$f_T(T)=[1-f_0(T)]$ are indicated on the right. We observe that the
condensate fractions are unity at $T=0$, corresponding to a completely
coherent boson system at $T=0$.  It decreases slowly as the
temperature increases, and the rate of decrease is small at low
temperatures.  The greater the number of bosons $N$, the larger is the
range of temperatures in which the boson system contains a substantial
fraction of the condensate.  For example, for a system with 2000
identical bosons, substantial fraction of the condensate occurs up to
$T/\hbar \omega \sim 11$.  The transition from the condensate phase to
the chaotic phase occurs over a large range of temperatures and is
therefore not a sharp first-order type transition.  The complementary
chaotic fraction $f_T(T)$ increases gradually as the temperature
increases, reaching the value of unity at $T/\hbar \omega \sim 11$ for
$N=2000$.

As the number of particles $N$ decreases down to 250, a substantial
condensate fraction occurs only for $T/\hbar \omega < 5$.  The chaotic
fraction $f_T(T)$ increases as a function of temperature and it reaches
the value of unity for $T/\hbar \omega \sim 6$.  The transition from
the condensate phase to the chaotic phase occurs over a temperature
range from $T/\hbar \omega\sim 2$ to $T/\hbar \omega\sim 5$.  The
smaller the number of particles, the lower the condensate temperature
$T_c$ and the smaller is the range of temperatures over which the
condensate phase transition occurs.

In the transitional region below $T_c$ with a substantial fraction of the
condensate, one can get an approximate value of the condensate
fraction by noting that in this region, the value of $z$ is close to
unity (Fig. 1).  The number of chaotic particles $N_T$ can be
estimated from Eq.\ (\ref{eq5}) by setting $z$ to unity, and we obtain
\begin{eqnarray}
\label{eq8}
N_T \sim  \left ( \frac{T}{\hbar \omega} \right )^3\sum_{k=1}^\infty
e^{-k\hbar \omega/2T}\left [ \frac {1}{k^3} +\frac {2(\hbar\omega/T) }{k^2}
+\frac{15(\hbar\omega/T)^2}{8k} \right ].
\end{eqnarray}
Consequently one can fit the condensate fraction $f_0(T)$ reasonably
well by a one-parameter function of the form
\begin{eqnarray}
\label{fit}
f_0(T)=1-(T/T_c)^3  {\rm ~~~~~ for~~} T \le T_c,
\end{eqnarray}
\begin{eqnarray}
f_0(T)=O(1/N)\to 0   {\rm ~~~~~ for~~} T \ge T_c.
\end{eqnarray}
The results from the one-parameter fit to $f_0(T)$ are shown as the
dashed curves in Fig.\ 2, to be compared with the $f_0(T)$ calculated
with the condensate configuration condition Eq.\ (\ref{eq6}) shown as
the solid curves.  The values of $T_c/\hbar \omega$ that give the best
fit to $f_0(T)$ for different $N$ values are listed in Table I.

The $T_c$ values can also be determined approximately by considering
the case of $\hbar \omega/T <<1$ in Eq.\ (\ref{eq8}) and we have
\begin{eqnarray}
N_T \sim  \left ( \frac{T}{\hbar \omega} \right )^3\zeta(3),
\end{eqnarray}
where $\zeta(3)=\sum_{k=1}^\infty k^{-3}=1.202$ is the zeta function
with the argument 3.  Thus, the condensate fraction is given approximately by
\begin{eqnarray}
\label{f00}
f_0(T)\sim 1-({T}/{T_{c,approx}})^3  {\rm ~~~~~for~~} T<T_{c,approx},
\end{eqnarray}
with 
\begin{eqnarray}
\label{tc}
\frac {T_{c, approx}}{\hbar \omega} 
\sim \left (\frac{N}{\zeta (3)} \right )^{1/3}=
\left (\frac{N}{1.202} \right )^{1/3}.
\end{eqnarray}

A comparison of the above approximate result with $T_c$ in Table I
indicates that the above Eqs. (\ref{f00}) and (\ref{tc}) are
approximately valid, with the values of $T_{c,{\rm approx}}$
determined by Eq.\ (\ref{tc}) slightly greater than $T_c$ by about
10\%.

\vskip 0.4cm \centerline{Table I.  Condensation
temperature $T_c/\hbar \omega$ of Eq. (\ref{fit})}

\centerline{ and $T_{c,approx}/\hbar\omega$ of Eq.\ (\ref{tc}) as a function
of $N$} {\vskip 0.3cm\hskip 2.0cm
\begin{tabular}{|c|c|c|c|}
\hline
{\rm Number of Bosons }$N$ & ~~~$T_c/\hbar \omega$~~~
                 &~ $T_{c,approx}/\hbar \omega=(N/1.202)^{1/3}$
                 &  $T_{c,approx}/T_c$\\
\hline
       2000      &  10.97
                 &  11.85 
                 &   1.08
\\ \cline{1-4}  
       1000      &   8.56
                 &   9.41 
                 &   1.10
\\ \cline{1-4}  
        500      &   6.63
                 &   7.47 
                 &   1.13
\\ \cline{1-4}  
        250      &   5.12
                 &   5.92 
                 &   1.16
\\ \cline{1-4}  
\end{tabular}
}
\vspace*{0.3cm}

\section{One-Body and Two-Body Density Matrices in Momentum Space}

Previously, the one- and two-body density matrices have been obtained
for identical bosons in configuration space \cite{Nar99}.  We would
like to write down the corresponding one-body and two-body density
matrices in momentum space so as to evaluate the momentum correlation
function. The results in momentum space can be readily obtained
from the results in configuration space by replacing $\bbox{r}$ in
Ref. \cite{Nar99} with $\bbox{p}$.  We thus have the one-body density
matrix in momentum space
\begin{eqnarray}
\label{one}
G^{(1)}(\bbox{p}_1,\bbox{p}_2)= \sum_{\rm n} 
u_{\rm n}^*(\bbox{p}_1)
u_{\rm n}  (\bbox{p}_2) \langle \hat a_{\rm n}^\dagger \hat a_{\rm n} \rangle.
\end{eqnarray}
Similarly, we have the two-body density matrix in momentum space
 given by
\begin{eqnarray}
G^{(2)}(\bbox{p}_1,\bbox{p}_2; \bbox{p}_1,\bbox{p}_2) 
=
\sum_{\rm klmn}
u_{\rm k}^*(\bbox{p}_1) u_{\rm l}^*(\bbox{p}_2)
u_{\rm m}(\bbox{p}_2) u_{\rm n} (\bbox{p}_1)
 \langle \hat a_{\rm k}^\dagger \hat a_{\rm l}^\dagger 
\hat a_{\rm m} \hat a_{\rm n}
\rangle.
\end{eqnarray}
We shall follow \cite{Nar99} in expressing the two-body density matrix
in terms of one-body density matrices.  By separating out the term
with ${\rm k}={\rm l}={\rm m}={\rm n}$ from other terms and using the
definition of the one-body density matrix (\ref{one}), the two-body
density matrix can be shown to be
\begin{eqnarray}
\label{eq10}
G^{(2)}(\bbox{p}_1,\bbox{p}_2; \bbox{p}_1,\bbox{p}_2) 
&=&G^{(1)}(\bbox{p}_1,\bbox{p}_1) G^{(1)}(\bbox{p}_2,\bbox{p}_2)
+ | G^{(1)}(\bbox{p}_1,\bbox{p}_2)|^2
\nonumber\\
&+&\sum_{\rm n=0}^{\infty}
|u_{\rm n}^*(\bbox{p}_1)|^2 | u_{\rm n}(\bbox{p}_2)|^2 
 \left \{ \langle \hat a_{\rm n}^\dagger 
\hat a_{\rm n}^\dagger \hat a_{\rm n} \hat a_{\rm n}
\rangle
- 2  
\langle \hat a_{\rm n}^\dagger \hat a_{\rm n}
\rangle
\langle \hat a_{\rm n}^\dagger \hat a_{\rm n}
\rangle \right \}.
\end{eqnarray}
The last term in this equation involves a summation over the
${\rm n}=0$ condensate state and the set of $\{{\rm n}>0\}$ states.
In line with our earlier discussions on the statistical ensemble for
the states \cite{Pol95,Nar99}, we shall use the grand canonical
ensemble for the set of $\{ {\rm n}>0\} $ states and the canonical
ensemble for the condensate state of ${\rm n}=0$.  For the set of
$\{{\rm n}>0\}$ states in the grand canonical ensemble then, the
occupation fluctuation characteristics of the grand canonical
ensemble make the contributions of the set of $\{{\rm n}>0\}$ states
small in comparison with the other terms on the right-hand side of the
above equation, as we shall see from the following discussion.  We
note that in the above equation
\begin{eqnarray}
\langle \hat a_{\rm n}^\dagger 
\hat a_{\rm n}^\dagger \hat a_{\rm n} \hat a_{\rm n}
\rangle
- 2  
\langle \hat a_{\rm n}^\dagger \hat a_{\rm n}
\rangle
\langle \hat a_{\rm n}^\dagger \hat a_{\rm n}
\rangle 
=\langle ( \hat a_{\rm n}^\dagger \hat a_{\rm n}
   -\langle\hat a_{\rm n}^\dagger \hat a_{\rm n}\rangle )^2 \rangle 
-\langle \hat a_{\rm n}^\dagger \hat a_{\rm n}\rangle
 \langle \hat a_{\rm n}^\dagger \hat a_{\rm n} \rangle .
\end{eqnarray}
For an ${\rm n}>0$ state in the grand canonical ensemble, the
mean-square fluctuation of the occupation number $\langle \hat a_{\rm
n}^\dagger \hat a_{\rm n}\rangle$ in the state is given by
\cite{Kit80}
\begin{eqnarray}
\langle ( \hat a_{\rm n}^\dagger \hat a_{\rm n}
   -\langle\hat a_{\rm n}^\dagger \hat a_{\rm n}\rangle )^2 \rangle 
=\langle\hat a_{\rm n}^\dagger \hat a_{\rm n}\rangle
\left ( \langle\hat a_{\rm n}^\dagger \hat a_{\rm n}\rangle +1
\right ).
\end{eqnarray}
Therefore, for this ${\rm n}>0$ state in the grand canonical ensemble,
we have 
\begin{eqnarray}
\langle \hat a_{\rm n}^\dagger 
\hat a_{\rm n}^\dagger \hat a_{\rm n} \hat a_{\rm n}
\rangle
- 2  
\langle \hat a_{\rm n}^\dagger \hat a_{\rm n}
\rangle
\langle \hat a_{\rm n}^\dagger \hat a_{\rm n}
\rangle
= \langle \hat a_{\rm n}^\dagger \hat a_{\rm n} \rangle,
\end{eqnarray}
and the contribution of the set of $\{{\rm n}>0\}$ states to the
two-body density matrix is
\begin{eqnarray}
\sum_{{\rm n}>0} |u_{\rm n}^*(\bbox{p}_1)|^2 | u_{\rm
  n}(\bbox{p}_2)|^2\langle \hat a_{\rm n}^\dagger \hat a_{\rm n}
\rangle.
\end{eqnarray}
When we integrate over $\bbox{p}_1$ and $\bbox{p}_2$, the set of
$\{ {\rm n}>0\}$ states gives
a contribution of 
\begin{eqnarray}
\int d\bbox{p}_1 ~d\bbox{p}_2 \sum_{{\rm n}>0} 
 |u_{\rm n}^*(\bbox{p}_1)|^2 | u_{\rm n}
 (\bbox{p}_2)|^2\langle \hat a_{\rm n}^\dagger \hat a_{\rm n}
\rangle = N_T,
\end{eqnarray}
whereas the other terms such as the first term of Eq.\ (\ref{eq10}),
$G^{(1)}(\bbox{p}_1,\bbox{p}_1) G^{(1)}(\bbox{p}_2,\bbox{p}_2)$, give
a contribution of $N^2$.  The contribution from the set of $\{ {\rm
n}>0\}$ states is $N_T/N^2$ of the contribution from
$G^{(1)}(\bbox{p}_1,\bbox{p}_1) G^{(1)}(\bbox{p}_2,\bbox{p}_2)$.
Therefore, in the limit of a large number of bosons $N$, the ratio
$N_T/N^2$ is small, and the contributions from the set of $\{{\rm
n}>0\}$ states in the summation in Eq.\ (\ref{eq10}) can be
neglected. We are left with only the ${\rm n}=0$ condensate state
contribution for this summation.

To describe the ${\rm n}=0$ condensate state, we shall follow Ref.\
\cite{Pol95,Nar99} and use the canonical ensemble which gives the
canonical fluctuation \cite{Pol95}
\begin{eqnarray}
\langle ( \hat a_{\rm n}^\dagger \hat a_{\rm n}
   -\langle\hat a_{\rm n}^\dagger \hat a_{\rm n}\rangle )^2 \rangle 
= \langle \hat a_0^\dagger \hat a_0^\dagger \hat a_0 \hat a_0
\rangle- \langle \hat a_0^\dagger \hat a_0\rangle  
  \langle \hat a_0^\dagger \hat a_0\rangle
=O(N_0).
\end{eqnarray}
Thus, we have
\begin{eqnarray}
\langle \hat a_{\rm n}^\dagger 
\hat a_{\rm n}^\dagger \hat a_{\rm n} \hat a_{\rm n}
\rangle
- 2  
\langle \hat a_{\rm n}^\dagger \hat a_{\rm n}
\rangle
\langle \hat a_{\rm n}^\dagger \hat a_{\rm n}
\rangle 
= - \langle \hat a_0^\dagger \hat a_0\rangle  
  \langle \hat a_0^\dagger \hat a_0\rangle + O(N_0).
\end{eqnarray}
In the limit of a large number of particles, we can neglect the last
term $O(N_0)$ in the above equation which is small compared to the
first term of order $N_0^2$.  The two-body momentum density matrix of
Eq.\ (\ref{eq10}) is therefore
\begin{eqnarray}
\label{eq18}
G^{(2)}(\bbox{p}_1,\bbox{p}_2; \bbox{p}_1,\bbox{p}_2) 
=G^{(1)}(\bbox{p}_1,\bbox{p}_1) G^{(1)}(\bbox{p}_2,\bbox{p}_2)
+ | G^{(1)}(\bbox{p}_1,\bbox{p}_2)|^2
-N_0^2 
|u_0(\bbox{p}_1)|^2 | u_0(\bbox{p}_2)|^2, 
\end{eqnarray}
which gives the conditional probability for the occurrence of a pion
of momentum $\bbox{p}_1$ in coincidence with another identical pion of
momentum $\bbox{p}_2$.  This two-body density matrix in momentum
space has the same form as that obtained earlier in configuration
space in \cite{Nar99}.

\section{The Momentum Correlation Function}

In Hanbury-Brown-Twiss measurements, we normalize the probability
relative to the probability of detecting particle $\bbox{p}_1$ and
$\bbox{p}_2$, and define the momentum correlation function
$C(\bbox{p}_1,\bbox{p}_2)$ as
\begin{eqnarray}
C(\bbox{p}_1,\bbox{p}_2)=
\frac {G^{(2)}(\bbox{p}_1,\bbox{p}_2; \bbox{p}_1,\bbox{p}_2)}
      {G^{(1)}(\bbox{p}_1,\bbox{p}_1) G^{(1)}(\bbox{p}_2,\bbox{p}_2)}.
\end{eqnarray}
It is convenient to introduce the average and the relative momenta 
of the pair
\begin{eqnarray}
\bbox{p}=(\bbox{p}_1+\bbox{p}_2)/2,
~~~~~~~~\bbox{q}=\bbox{p}_1-\bbox{p}_2,
\end{eqnarray}
with the inverse transformation
\begin{eqnarray}
\bbox{p}_1=\bbox{p}+\frac{\bbox{q}}{2}, 
~~~~~~~~
\bbox{p}_2=\bbox{p}-\frac{\bbox{q}}{2}. 
\end{eqnarray}
The momentum correlation function can be expressed alternatively in
terms of the kinematic variables $\bbox{p}$ and $\bbox{q}$.  From
Eq. (\ref{eq18}), we have the general expression for the correlation
function
\begin{eqnarray}
\label{eq22}
C(\bbox{p},\bbox{q})=C(\bbox{p}_1,\bbox{p}_2)=1+
\frac {|G^{(1)}(\bbox{p}_1,\bbox{p}_2)|^2-N_0^2 |u_0(\bbox{p}_1)|^2
                                                |u_0(\bbox{p}_2)|^2}
      {G^{(1)}(\bbox{p}_1,\bbox{p}_1) G^{(1)}(\bbox{p}_2,\bbox{p}_2)}.
\end{eqnarray}
In the nearly completely coherent case with almost all particles in
the ground condensate state, $N_0 \to N$, the two terms in the
numerator cancel each other and we have $C(p,q)=1$, as it should
be.  For the other extreme of a completely chaotic source with
$N_0<<N$, the second term in the numerator proportional to $N_0^2$
gives negligible contribution and can be neglected.  The correlation
function becomes the usual one for a completely chaotic source,
\begin{eqnarray}
C_{\rm chaotic}(\bbox{p},\bbox{q}) =1+
\frac {|G^{(1)}(\bbox{p}_1,\bbox{p}_2)|^2} 
      {G^{(1)}(\bbox{p}_1,\bbox{p}_1) G^{(1)}(\bbox{p}_2,\bbox{p}_2)}.
\end{eqnarray}
The general result of Eq. (\ref{eq22}) allows one to study the
correlation function for all cases with varying degrees of coherence.

If we introduce $R(\bbox{p},\bbox{q})= R(\bbox{p}_1,\bbox{p}_1)=
C(\bbox{p},\bbox{q})-1$, then
\begin{eqnarray}
R(\bbox{p},\bbox{q})=
R(\bbox{p}_1,\bbox{p}_2)=
\frac {|G^{(1)}(\bbox{p}_1,\bbox{p}_2)|^2-N_0^2 |u_0(\bbox{p}_1)|^2
                                                |u_0(\bbox{p}_2)|^2}
      {G^{(1)}(\bbox{p}_1,\bbox{p}_1) G^{(1)}(\bbox{p}_2,\bbox{p}_2)}.
\end{eqnarray}

It is of interest to express the momentum correlation function
$C(p,q)$ in terms of the Wigner function $f(r,p)$ defined in terms of
the the one-body density matrix $G^{(1)}(\bbox{r}_1,\bbox{r}_2) $ as
\begin{eqnarray}
f(\bbox{r},\bbox{p})=
\int d\bbox{s} ~e^{i \bbox{p}\cdot \bbox{s}}
G^{(1)}(\bbox{r}+\frac{\bbox{s}}{2},\bbox{r}-\frac{\bbox{s}}{2}),
\end{eqnarray}
This one-body density matrix in configurations space is related to the
one-body density matrix in momentum space by a Fourier transform,
\begin{eqnarray}
G^{(1)}(\bbox{p}_1,\bbox{p}_2)
=\int d\bbox{r}_1~ d\bbox{r}_2
~e^{i \bbox{p}_1 \cdot \bbox{r}_1 - i \bbox{p}_2 \cdot \bbox{r}_2}
G^{(1)}(\bbox{r}_1,\bbox{r}_2).
\end{eqnarray}
Therefore, by changing coordinates from $\bbox{r}_1$ and $\bbox{r}_2$
to $\bbox{r}=(\bbox{r}_1+\bbox{r}_2)/2$ and
$\bbox{s}=\bbox{r}_1-\bbox{r}_2$, we can relate the one-body density
$G^{(1)}(\bbox{p}_1,\bbox{p}_2)$ with the Wigner function
$f(\bbox{r},\bbox{p})$,
\begin{eqnarray}
G^{(1)}(\bbox{p}_1,\bbox{p}_2)
=\int d\bbox{r} e^{i \bbox{q}\cdot \bbox{r}}~ f(\bbox{r},\bbox{p}),
\end{eqnarray}
and in particular, for the diagonal density matrix element we have
\begin{eqnarray}
G^{(1)}(\bbox{p}_1,\bbox{p}_1)
=\int d\bbox{r} f(\bbox{r},\bbox{p}_1).
\end{eqnarray}
As a consequence, the correlation function $C(p,q)$ in
Eq. (\ref{eq22}) can be rewritten as 
\begin{eqnarray}
\label{eq14p}
C(\bbox{p},\bbox{q})=1+
\frac {|\int d\bbox{r} e^{i \bbox{q}\cdot \bbox{r}}
~ f(\bbox{r},\bbox{p})|^2-N_0^2 
|u_0(\bbox{p}+\bbox{q}/2)|^2 |u_0(\bbox{p}-\bbox{q}/2)|^2}
      { \int d\bbox{r} f(\bbox{r},\bbox{p}+\bbox{q}/2) 
        \int d\bbox{r} f(\bbox{r},\bbox{p}-\bbox{q}/2)}.
\end{eqnarray}
This is the general expression for the momentum correlation function
expressed in terms of the Wigner function $f(\bbox{r},\bbox{p})$ when
the coherence of the system is properly taken into account.

The $R$ function for the general case is related to the Wigner
function $f(\bbox{r},\bbox{p})$ by,
\begin{eqnarray}
\label{eq14q}
R(\bbox{p},\bbox{q})=
\frac {|\int d\bbox{r} e^{i \bbox{q}\cdot \bbox{r}}
~ f(\bbox{r},\bbox{p})|^2-N_0^2 
|u_0(\bbox{p}+\bbox{q}/2)|^2 |u_0(\bbox{p}-\bbox{q}/2)|^2}
      { \int d\bbox{r} f(\bbox{r},\bbox{p}+\bbox{q}/2) 
\int d\bbox{r} f(\bbox{r},\bbox{p}-\bbox{q}/2)}.
\end{eqnarray}

When the condensate fraction $f_0$ is large with $N_0 \to N$, the
second term in the numerator of the above equations is important and
must be properly taken into account. In fact, in the completely
coherent case, Eqs.\ (\ref{eq14p}) and (\ref{eq14q}) give $C(p,q)=1$ and
$R(p,q)=0$.  Only in the special case of a completely chaotic state is
the contribution from the second term in the numerator negligible, and
we have the usual relationship between the Wigner function and the
momentum correlation function for a chaotic system,
\begin{eqnarray}
C_{\rm chaotic} (\bbox{p},\bbox{q})\sim 1+
\frac {|\int d\bbox{r} e^{i \bbox{q}\cdot \bbox{r}}~ f(\bbox{r},\bbox{p})|^2}
      { \int d\bbox{r} f(\bbox{r},\bbox{p}+\bbox{q}/2) \int d\bbox{r} f(\bbox{r},\bbox{p}-\bbox{q}/2)}.
\end{eqnarray}

\section{The One-Body Density Matrix for a Harmonic Oscillator Potential}

For a given total particle number $N$ of particle mass $m$ in an
external harmonic oscillator potential, we have obtained in Section II
the fugacity $z$ as a function of $T/\hbar \omega$ (Fig.\ 1). This
solution of $z$ allows us to evaluate the density matrices and the
correlation functions at various temperatures.  For the harmonic
oscillator potential, the one-body density matrix has been obtained
previously in configuration space as given by \cite{Nar99}
\begin{eqnarray}
G^{(1)}(\bbox{r}_1,\bbox{r}_2) 
&=&\sum_{{\rm n}=0}^\infty 
u_{\rm n}^*(\bbox{r}_1)
u_{\rm n}  (\bbox{r}_2) \frac{z e^{-\beta \tilde\epsilon_{\rm n}}}
                   {1- z e^{-\beta \tilde\epsilon_{\rm n}}}
\nonumber \\
&=&\sum_{k=1}^\infty z^k 
{\tilde G}_0 (\bbox{r}_1,\bbox{r}_2; k \beta \hbar \omega),
\end{eqnarray}
where
\begin{eqnarray}
{\tilde G}_0 (\bbox{r}_1,\bbox{r}_2; \tau)
= \left (  \frac {1}{\pi a^2 (1-e^{-2\tau})} \right )^{3/2}
\exp\left ( -\frac {1}{a^2} 
\frac {(\bbox{r}_1^2 + \bbox{r}_2^2)(\cosh~\tau -1)+(\bbox{r}_1-\bbox{r}_2)^2}
      {2\sinh \tau}  \right ). 
\end{eqnarray}
Because of the exchange symmetry of $\bbox{r}/a$ and $\bbox{p}a/\hbar$
for a harmonic oscillator potential, the one-body density matrix in
momentum space can be readily obtained from these results of
Ref. \cite{Nar99} by replacing $\bbox{r}/a$ with $\bbox{p}a/\hbar$,
and we get
\begin{eqnarray}
G^{(1)}(\bbox{p}_1,\bbox{p}_2) 
&=&\sum_{{\rm n}=0}^\infty 
u_{\rm n}^*(\bbox{p}_1)
u_{\rm n}  (\bbox{p}_2) \frac{z e^{-\beta \tilde\epsilon_{\rm n}}}
                   {1- z e^{-\beta \tilde\epsilon_{\rm n}}}
\nonumber \\
&=&\sum_{k=1}^\infty z^k 
{\tilde G}_0 (\bbox{p}_1,\bbox{p}_2; k \beta \hbar \omega),
\end{eqnarray}
where
\begin{eqnarray}
{\tilde G}_0 (\bbox{p}_1,\bbox{p}_2; \tau)
= \left (  \frac {a^2}{\pi \hbar^2 (1-e^{-2\tau})} \right )^{3/2}
\exp\left ( -\frac {a^2}{\hbar^2} 
\frac {(\bbox{p}_1^2 + \bbox{p}_2^2)(\cosh~\tau -1)+(\bbox{p}_1-\bbox{p}_2)^2}
      {2\sinh \tau}  \right ). 
\end{eqnarray}
We can write ${\tilde G}_0 (\bbox{p}_1,\bbox{p}_2; \tau)$ in terms of
the ground state wave function $u_0^*(\bbox{p}_1)u_0(\bbox{p}_2)$ as
\begin{eqnarray}
{\tilde G}_0 (\bbox{p}_1,\bbox{p}_2; \tau) 
=u_0^*(\bbox{p}_1) u_0(\bbox{p}_2){\tilde g}_0(\bbox{p}_1,\bbox{p}_2;\tau) 
\end{eqnarray}
where the ground state wave function is
\begin{eqnarray}
u_0(\bbox{p})=\left (\frac{a^2}{\pi \hbar^2}\right )^{3/4} 
\exp \left \{ -\frac{a^2}{\hbar^2}\frac{\bbox{p}^2}{2} \right \},
\end{eqnarray}
and the dimensionless function ${\tilde g}_0
(\bbox{p}_1,\bbox{p}_2;\tau)$ is given by
\begin{eqnarray}
{\tilde g}_0 (\bbox{p}_1,\bbox{p}_2;\tau) = \frac
{1}{(1-e^{-2\tau})^{3/2}} \exp\left ( -\frac {a^2}{\hbar^2} \frac
{(\bbox{p}_1^2 + \bbox{p}_2^2)(\cosh~\tau -1-\sinh \tau )
+(\bbox{p}_1-\bbox{p}_2)^2} {2\sinh \tau} \right ).
\end{eqnarray}
Then we have
\begin{eqnarray}
\label{g1}
G^{(1)}(\bbox{p}_1,\bbox{p}_2)
=u_0^*(\bbox{p}_1) u_0(\bbox{p}_2) A(\bbox{p}_1,\bbox{p}_2) 
\end{eqnarray}
where
\begin{eqnarray}
\label{bigA}
A(\bbox{p}_1,\bbox{p}_2)=\sum_{k=1}^\infty z^k
{\tilde g}_0 (\bbox{p}_1,\bbox{p}_1;k \beta \hbar \omega).
\end{eqnarray}
In numerical calculations, especially at low temperatures where $z$ is
close to unity, the number of terms in the summation over $k$ in
$A(\bbox{p}_1,\bbox{p}_2)$ will need to be greater than the number of
particles $N$.  To avoid such a lengthy summation, it is simplest to
separate out the condensate component to write the above as
\begin{eqnarray}
\label{bigAA}
A(\bbox{p}_1,\bbox{p}_2)=\frac{z}{1-z}
+\sum_{k=1}^\infty z^k 
[{\tilde g}_0 (\bbox{p}_1,\bbox{p}_1; k \beta \hbar \omega)-1].
\end{eqnarray}
For low temperatures, the coefficient $[{\tilde
g}_0(\bbox{p}_1,\bbox{p}_1; k \beta \hbar \omega)-1]$ of $z^k$ is
small and a small number of terms in $k$ will suffice.  For high
temperatures above the condensate temperature, $z$ is substantially
less than unity, and $z^k$ decreases rapidly as $k$ increases; a small
number of terms in $k$ will also suffice.

From Eq.\ (\ref{eq22}) the momentum correlation function is
\begin{eqnarray}
\label{cq}
C(\bbox{p},\bbox{q})
=C(\bbox{p}_1,\bbox{p}_2)=1+
\frac {|A(\bbox{p}_1,\bbox{p}_2)|^2 - |z/(1-z)|^2}
      { A(\bbox{p}_1,\bbox{p}_1) A(\bbox{p}_2,\bbox{p}_2)}.
\end{eqnarray}
and
\begin{eqnarray}
R(\bbox{p},\bbox{q})
=R(\bbox{p}_1,\bbox{p}_2)=
\frac {|A(\bbox{p}_1,\bbox{p}_2)|^2 - |z/(1-z)|^2}
      { A(\bbox{p}_1,\bbox{p}_1) A(\bbox{p}_2,\bbox{p}_2)}.
\end{eqnarray}

The one-body Wigner function for the boson system can be obtained from
the one-body density matrix $G^{(1)}(\bbox{r}_1,\bbox{r}_2) $ and we
find
\begin{eqnarray}
f(\bbox{r},\bbox{p})=
\sum_{k=1}^\infty z^k 
\left (  \frac {4 \tanh(k\beta\hbar \omega/2)}
               {(1-e^{-2k\beta\hbar\omega})} \right )^{3/2}
\exp\left \{ - \left ( \frac {\bbox{r}^2}{a^2}+\frac{\bbox{p}^2 a^2}{\hbar^2}\right ) 
      \tanh \left (\frac{k\beta\hbar\omega}{2}\right )  \right \}.
\end{eqnarray}
There is an explicit symmetry between $\bbox{x}/a$ and $\bbox{p}a$ in
the Wigner function for the harmonic oscillator potential.

\section{Spatial and momentum distributions}

\begin{figure}[h]
\includegraphics[scale=0.5]{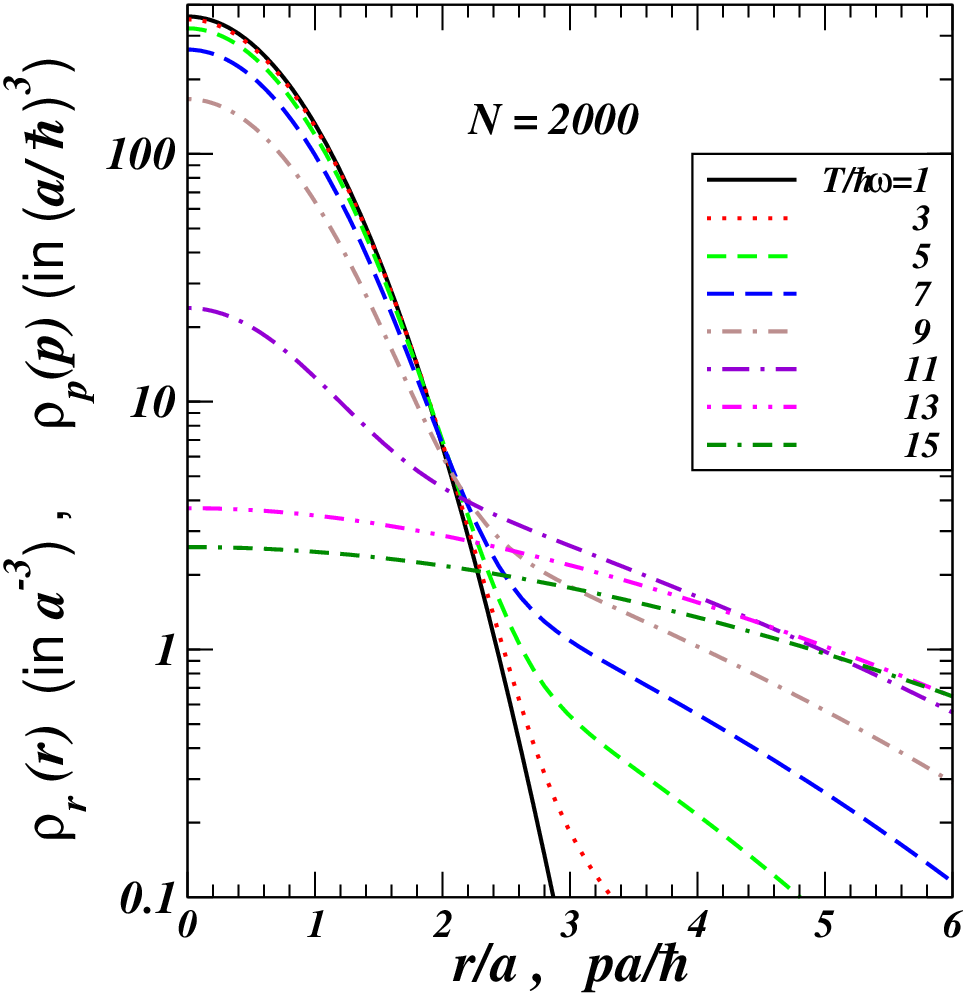}
\caption{(Color online) The spatial density distribution
$\rho_r(\bbox{r})$ in units of $a^{-3}$, expressed as a function of
$r/a$, and the momentum density distribution $\rho_p(\bbox{p})$ in
units of $(a/\hbar)^3$, expressed as a function of $pa/\hbar$, for a
boson system with $N=2000$ at different temperatures.  }
\end{figure}

Before we evaluate the momentum correlation function $C(p,q)$, it is
useful to study the single-particle spatial and momentum distributions
$\rho_r (\bbox{r})$ and $\rho_p (\bbox{p})$.  Because of the symmetry
between $\bbox{r}/a$ and $\bbox{p}a/\hbar$ in a harmonic oscillator
potential, the following two functions have the same shape: (i)
$\rho_r(\bbox{r})$ in units of $a^{-3}$ expressed as a function of
$\bbox{r}/a$, and (ii) $\rho_p(\bbox{p})$ in units of $(a/\hbar)^{3}$
expressed as a function of $\bbox{p}a/\hbar$.  The two distributions
can be displayed on the same graph.  From the one-body density matrix
(\ref{g1}), we obtain
\begin{eqnarray}
\rho_p(\bbox{p})=G^{(1)}(\bbox{p},\bbox{p})
=\left (\frac{a^2}{\pi \hbar^2}\right )^{3/2} 
\exp \left \{ -\frac{a^2\bbox{p}^2}{\hbar^2} \right \} A(\bbox{p},\bbox{p}), 
\end{eqnarray}
where
\begin{eqnarray}
A(\bbox{p},\bbox{p})=\sum_{k=1}^\infty z^k
{\tilde g}_0 (\bbox{p},\bbox{p};k \beta \hbar \omega),
\end{eqnarray}
\begin{eqnarray}
{\tilde g}_0 (\bbox{p},\bbox{p};\tau) = \frac
{1}{(1-e^{-2\tau})^{3/2}} \exp\left ( -\frac {a^2}{\hbar^2} \frac
{\bbox{p}^2(\cosh~\tau -1-\sinh \tau )} {\sinh \tau} \right ).
\end{eqnarray}
We plot in Fig. 3 the spatial and momentum distributions of the system
with $N=2000$ as a function of their dimensionless variables ${r}/a$
and ${p}a/\hbar$, respectively.  One observes that up to
$T/\hbar\omega\sim 9$ the system has a small spatial or momentum size
and there is a substantial condensate fraction in the system.  In
Fig. 4 we plot the root-mean-squared radius in units of $a$, $r_{\rm
rms}/a=\sqrt{\langle (r/a)^2\rangle}$, and the root-mean-squared
momentum in units of $\hbar/a$, $p_{\rm rms}a/\hbar=\sqrt{\langle
(pa/\hbar)^2\rangle}$, as a function of $T/\hbar \omega$.  For
$N=2000$, the quantity $r_{\rm rms}/a$ is slightly greater than 1 up
to $T/\hbar \omega \sim 6$, and it increases relatively rapidly to about
5.5 at the condensate temperature, $T_c/\hbar \omega \sim 11$.  It
increases at a relatively slower rate at temperatures above $T_c$. 

\begin{figure}[h]
\includegraphics[scale=0.5]{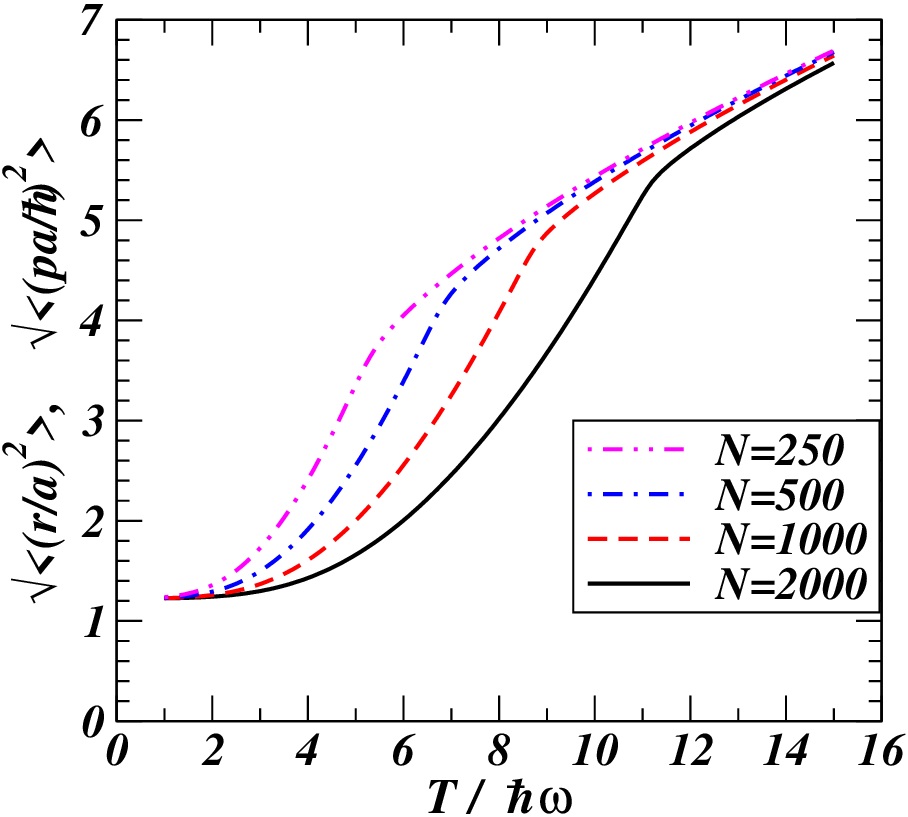}
\caption{(Color online) The root-mean-squared radius in unit of $a$
and the root-mean-squared momentum in units of $\hbar/a$, as a
function of $T/\hbar\omega$ for different numbers of bosons in the
system.}
\end{figure}

The size of the momentum distribution also undergoes similar changes
as a function of temperature.  The root-mean-squared momentum has the
dimension of about one unit of $\hbar/a$ at $T/\hbar \omega \sim 0$
and this linear size increases about sixfold when the temperature
reaches the chaotic region of $T_c/\hbar \omega \sim 11$ for $N=2000$.

We observe therefore that for a boson system in a harmonic oscillator,
the Bose-Einstein condensation gives rise to a distribution localized
in the region of small momentum and small spatial coordinates.  From
the viewpoints of the spatial and momentum densities, the
Bose-Einstein condensate in a harmonic oscillator is therefore a
``condensation'' in both momentum space and configuration space.

\section{Evaluation of the 
Momentum Correlation function $C(\bbox{p,q})$}

\begin{figure}[h]
\includegraphics[scale=0.5]{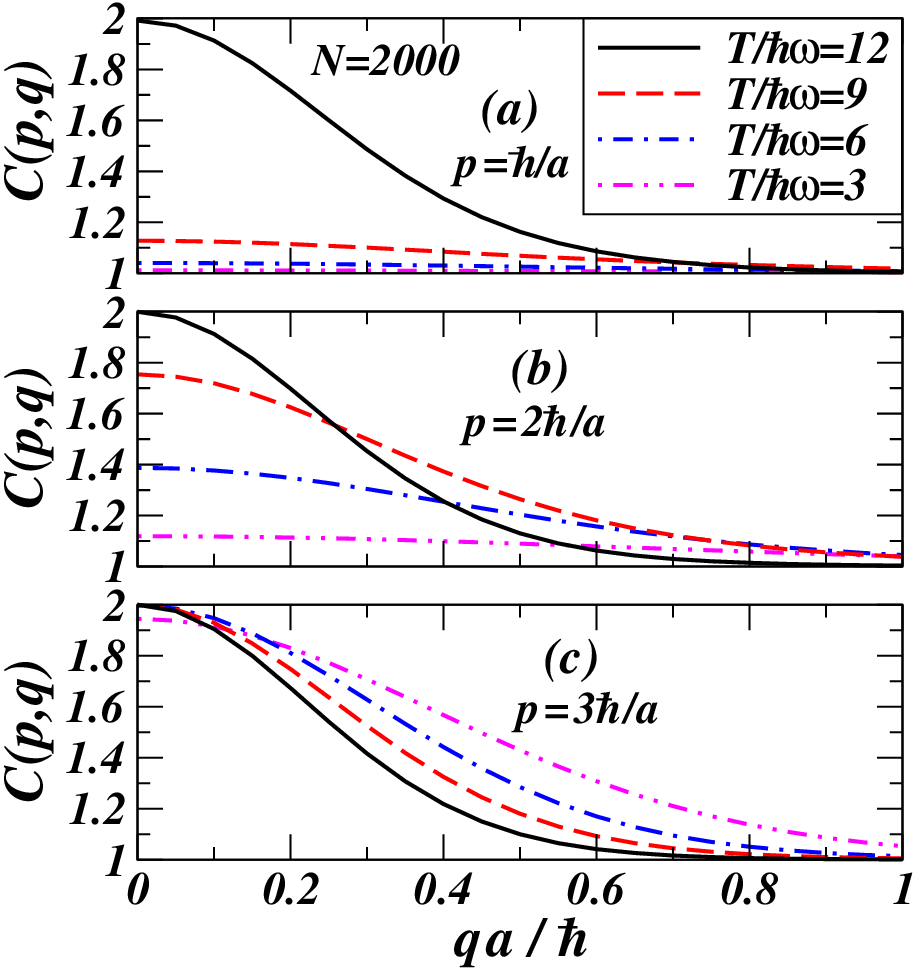}
\caption{(Color online) The correlation function at different values
of the pair momentum $pa/\hbar$ and temperatures.  Panels (a),
5(b), and (c) are for $p= 1,2,$ and 3 $\hbar/a$, respectively.  }
\end{figure}

With the solution $z$ obtained for a given $T/\hbar \omega$ as shown
in Fig.\ 1 and discussed in Section II, one can use Eq. (\ref{bigAA})
to evaluate $A(\bbox{p}_1,\bbox{p}_2)$.  The knowledge of
$A(\bbox{p}_1,\bbox{p}_2)$ then allows the determination of the
momentum correlation function from $C(p,q)$ using Eq. (\ref{cq}).

We show results of $C(p,q)$ for the case of $N=2000$ in Fig.\ 5.  We
observe that the correlation function is a complicated function of the
average pair momentum $p$ and temperature $T$.  For $p=\hbar/a$ in
Fig. 5($a$), the correlation function $C(p,q)$ at $q=0$ is close to
unity for temperatures below and up to $T/\hbar \omega=9$, but
increases to 2 rather abruptly at $ T/\hbar \omega=12$.  For
$p=2\hbar/a$ in Fig. 5($b$), the correlation function $C(p,q)$ at
$q=0$ is substantially above unity and increases gradually as
temperature increases.  For $p=3\hbar/a$ in Fig. 5($c$), the
correlation function $C(p,q)$ at $q=0$ is about 2 for all cases of
temperatures examined.

\begin{figure}[h]
\includegraphics[scale=0.5]{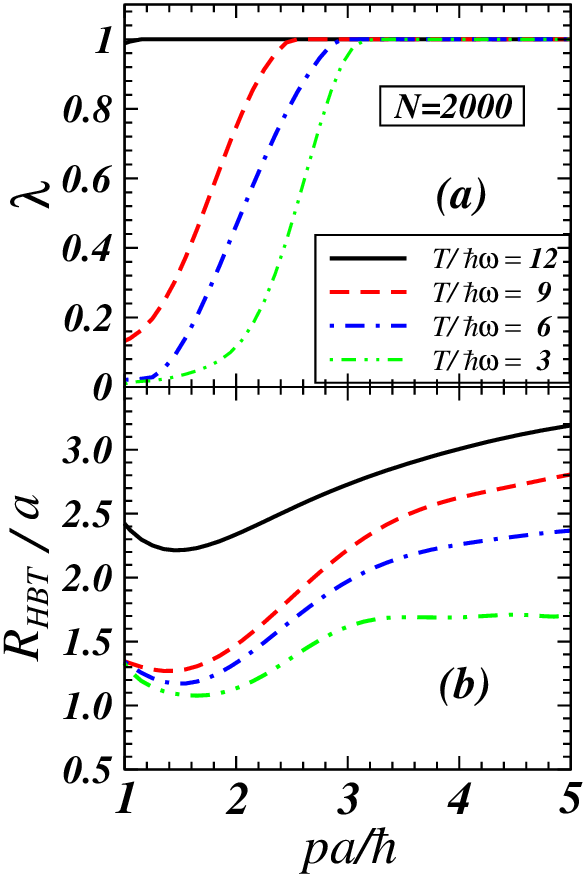}
\caption{(Color Online) ($a$) The parameter $\lambda$ as a function of $pa/\hbar$ for
different temperatures for $N=2000$.  ($b$) The ratio $R_{\rm HBT}/a$
as a function of $pa/\hbar$ for different temperatures for $N=2000$.
}
\end{figure}

If one follows the standard phenomenological analysis and introduces
the chaoticity parameter $\lambda$ to represent the correlation
function at zero relative momentum, then this parameter $\lambda$ is a
function of the average pair momentum $p$ and temperature $T$
\begin{eqnarray}
\lambda(p,T)=[C(p,q=0;T)-1],
\end{eqnarray}
where we display explicitly the dependence of the correlation function
on the temperature $T$.  We plot the values of $\lambda(p,T)$ as a
function of $p$ in Fig. 6($a$) for different temperatures for the case
of $N=2000$.  At $T/\hbar \omega=12$, which is above the condensate
temperature $T_c$, the $\lambda$ parameter is 1 for all $p$ values.
At $T/\hbar \omega=9$, which is below the condensate temperature
$T_c$, the $\lambda$ parameter drops precipitously to $\sim$0.1 at
$pa/\hbar=1$.  At this $T/\hbar \omega=9$, as $p$ increases the
$\lambda$ parameter rises gradually and reaches the constant value of
1 at $pa/\hbar=2.4$.  At $T/\hbar \omega=6$ and 3, for which the
systems are significantly coherent with large condensate fractions,
the $\lambda$ parameter starts close to zero at $pa/\hbar =1$, but as
$p$ increases the $\lambda$ parameter increases gradually to unity at
$pa/\hbar=2.9$ and 3.1 for $T/\hbar \omega=6$ and 3 respectively.  The
location where the $\lambda$ parameter attains unity changes with
temperature.  The lower the temperature, the greater is the value of
$p$ at which the $\lambda$ parameter attains unity.

We conclude from our results that the parameter $\lambda(p,T)$ is a
sensitive function of both $p$ and $T$ and $\lambda(p,T)=1$ is not a
consistent measure of the absence of the condensate fraction, as it
attains the value of unity in some kinematic regions for significantly
coherent systems with large condensate fractions at temperatures much
below $T_c$.  Only for the region of small $p$ will the parameter
$\lambda(p,T)$ be correlated with, but not equal to, the chaotic
fraction $f_T(T)$ of the system.

\begin{figure}[h]
\includegraphics[scale=0.5]{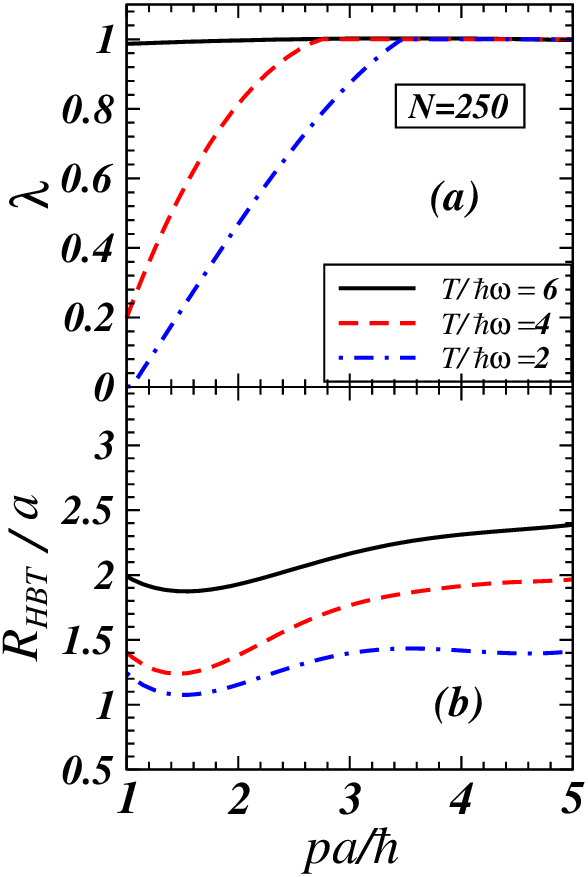}
\caption{(Color online) ($a$) The ratio $R_{\rm HBT}/a$ as a function
of $pa/\hbar$ for different temperatures for $N=250$.  ($b$) The
parameter $\lambda$ as a function of $pa/\hbar$ for different
temperatures for $N=250$.  }
\end{figure}

One can evaluate the root-mean-squared momentum in the correlation
function defined as
\begin{eqnarray}
q_{\rm rms}^2(p,T)=
\langle \bbox{q}^2\rangle = 
\frac{\int d\bbox{q} \bbox{q}^2 [C(p,q;T)-1]} {\int d\bbox{q} [C(p,q;T)-1]}
\end{eqnarray}
and introduce the HBT radius $R_{\rm HBT}(p,T)$ defined by
\begin{eqnarray}
R_{\rm HBT}(p,T)= \sqrt{\frac{3}{2}} \frac{\hbar}{q_{\rm rms}(p,T)}.
\end{eqnarray}
The HBT radius $R_{\rm HBT}(p,T)$ is in fact the radius parameter in
the standard Gaussian parametrization of the momentum correlation
function,
\begin{eqnarray}
C(p,q;T)=1+\lambda(p,T) \exp\{ - q^2 R_{\rm HBT}^2(p,T)/\hbar^2 \}.
\end{eqnarray}

We plot $R_{\rm HBT}(p,T)$ as a function of $pa/\hbar$ for different
temperatures $T$ for the case of $N=2000$ in Fig. 6($b$).  For fixed
values of temperature $T/\hbar \omega=3,6,$ and 9 and varying $p$,
one observes that $R_{\rm HBT}(p,T)$ is about $1.3a$ and $R_{\rm
HBT}(p,T)$ decreases slightly before it increases gradually as $p$
increases.  For $T/\hbar \omega=12$, which is above the condensate
temperature, $R_{\rm HBT}(p,T)$ is about $2.4a$ at $pa/\hbar=1$, and
$R_{\rm HBT}(p,T)$ decreases slightly before it increases gradually as
$p$ increases.  For a fixed value of $p$ at $pa/\hbar=1$, the HBT
radius increases with increasing temperatures very slowly at low
temperatures, and it increases rather abruptly when the condensation
temperature is approached.  For this small value of $pa/\hbar$, the
variation of $R_{\rm HBT}(p,T)$ as a function of $T$ reflects closely
the variation of the root-mean-squared radius as a function of
$T/\hbar \omega$, as shown in Fig. 4.  For large values of $p$,
$R_{\rm HBT}(p,T)$ increases with increasing temperatures in a more
uniform manner.

In Fig. 7($a$) and 7($b$), we show, respectively, $\lambda(p,T)$ and
$R_{\rm HBT}(p,T)$ as a function of $pa/\hbar$ and $T$ for the case
with $N=250$.  The $\lambda$ parameter and the HBT radius $R_{\rm
HBT}$ behave in a manner similar to those for the case of $N=2000$.
One observes in Fig. 7($a$) that at temperatures below $T_c$, the
$\lambda$ parameter is small at small $p$ and it increases as $p$
increases, reaching the saturating value of unity at $pa/\hbar =3.5$
for $T/\hbar \omega = 2$ and at $pa/\hbar =2.7$ for $T/\hbar \omega =
4$.  Above the condensation temperature at $T/\hbar \omega = 6$, the
$\lambda$ parameter assumes the value of unity for all $p$ values. 

As shown in Fig. 7($b$), for a fixed value of temperature $T$, the
HBT radius $R_{\rm HBT}$ decreases slightly and then increases
gradually with $p$.  For a fixed $p$ with a small $p$, the increase in
$R_{\rm HBT}$ is slow at low temperatures and the increase becomes
more rapid as the temperature approaches the condensate temperature of
$T_c/\hbar \omega=5.12$.

\section{Bose-Einstein Condensation of pions in a mean field 
(nonrelativistic)}

There is not much information on the magnitude of the mean-field
potential experienced by the pions. From the Glauber theory
\cite{Gla59}, the mean-field potential experienced by a pion in a pion
medium is related to the pion density $\rho_r ({\bf r})$ by
\begin{eqnarray}
V({\bf r})=-\frac{2\pi f(0)}{m} \rho_r ({\bf r}),
\end{eqnarray}
where $f(0)$ is the forward $\pi$-$\pi$ scattering amplitude.  We hope
to evaluate the pion mean-field potential in the future.  In the
meantime, the results in the previous sections allow us to answer the
following theoretical question.  If a system of $N$ pions is held
together by its mean field, taken to be a harmonic oscillator, and if
it comes to a state of static equilibrium with a given
root-mean-squared radius $r_{\rm rms}$ at a temperature $T$, what is
the condensate fraction of such a system?  The answer to this
theoretical question will provide useful information on the importance
of the Bose-Einstein condensation for a pion system in static
equilibrium, to pave the way for future investigations for the system
in dynamical expansion.

We would like to examine pion systems with a typical $r_{\rm rms}$,
$T$ and the pion number that one encounters in high-energy heavy-ion
collisions.  For a pion gas distribution with an HBT radius of about
$R_{_{\rm HBT}}=6$ fm as appropriate for Au-Au central collisions
\cite{Lis05}, the root-mean-squared radius $r_{\rm rms}$ for a Gaussian
density distribution is $\sqrt{3}R_{\rm HBT}$, which is about 10 fm.
We shall therefore examine a pion system with $r_{\rm rms}=10$ fm, a
temperature range from 80 to 160 MeV, and the number of pions to be
$N=250$ (for a central SPS Au-Au collision at $\sqrt{s_{NN}}=19.4$ GeV) and
$N=2000$ (for a central RHIC Au-Au collisions at $\sqrt{s_{NN}}=200$ GeV).

For a pion with a temperature of 80 to 160 MeV which is of the order
of the pion rest mass of 140 MeV, the motion of the pions is
relativistic and the proper treatment will need to be relativistic in
nature.  We shall carry out a relativistic treatment of the pion
states in the next section and shall content ourselves here in the
type of solution one gets in a nonrelativistic treatment.  Carrying
out both relativistic and nonrelativistic treatments will allow one
to understand what features of the coherence are sensitively affected
by the relativistic motion.

We first determine the strength of the mean-field potential $\hbar
\omega $ that can hold a system of $N$ pions in static equilibrium at
temperature $T$ for a given root-mean-squared radius of $r_{\rm rms}$.
For the pion system in static equilibrium, the quantity $r_{\rm
rms}/a$ is a function $F_N(x)$ of the variable $x=T/\hbar \omega $ as
shown in Fig. 4, where the subscript $N$ labels the boson number.  If
the value of $r_{\rm rms}$ is fixed as given, the quantities $\hbar
\omega $ and $T$ are then related by the set of parametric equations
\begin{eqnarray}
\label{pareq1}
\hbar \omega =\frac{[\hbar F_N(x)]^2}{r_{\rm rms}^2 m},
\end{eqnarray}
\begin{eqnarray}
\label{pareq2}
T=x \frac{[\hbar F_N(x)]^2}{r_{\rm rms}^2 m}.
\end{eqnarray}
By varying $x=T/\hbar \omega$ for a fixed $r_{\rm rms}$ and using the
function $F_N(x)$ of Fig. 4 in the above equations, the energy $\hbar
\omega $ can be determined as a function of $T$ for the cases of
$N=2000$ and $N=250$.  The results are shown in Fig. 8($a$).  One
finds that for the pion system with a given root-mean-squared radius of
10 fm, the value of $\hbar \omega $ ranges from about 12 to 20 MeV for
$N=2000$ and about 20 to 30 MeV for $N=250$. The ratio of
$T/\hbar\omega $ is about 7 for $N=2000$, and is about 4.5 for $N=250$,
as shown in Fig. 8($b$).  From these ratios of $T/\hbar\omega$, one
can use Fig. 2 to find out the condensate fraction.  The condensate
fractions $f_0(T)$ for a pion gas at various temperatures with
$N=2000$ and $N=250$ are shown in Fig. 8($c$).  One finds that
$f_0(T)$ is about $0.67-0.8$ for $N=2000$ and is about $0.9$ for
$N=250$.  The knowledge of $\hbar \omega$ in Fig. 8($a$) allows one to
determine the values of $a$ as a function of the temperature as shown
in Fig. 8($d$).  The oscillator length $a$ is about 4 fm for $N=2000$
and about 3.5 fm for $N=250$.

\begin{figure}[h]
\includegraphics[scale=0.5]{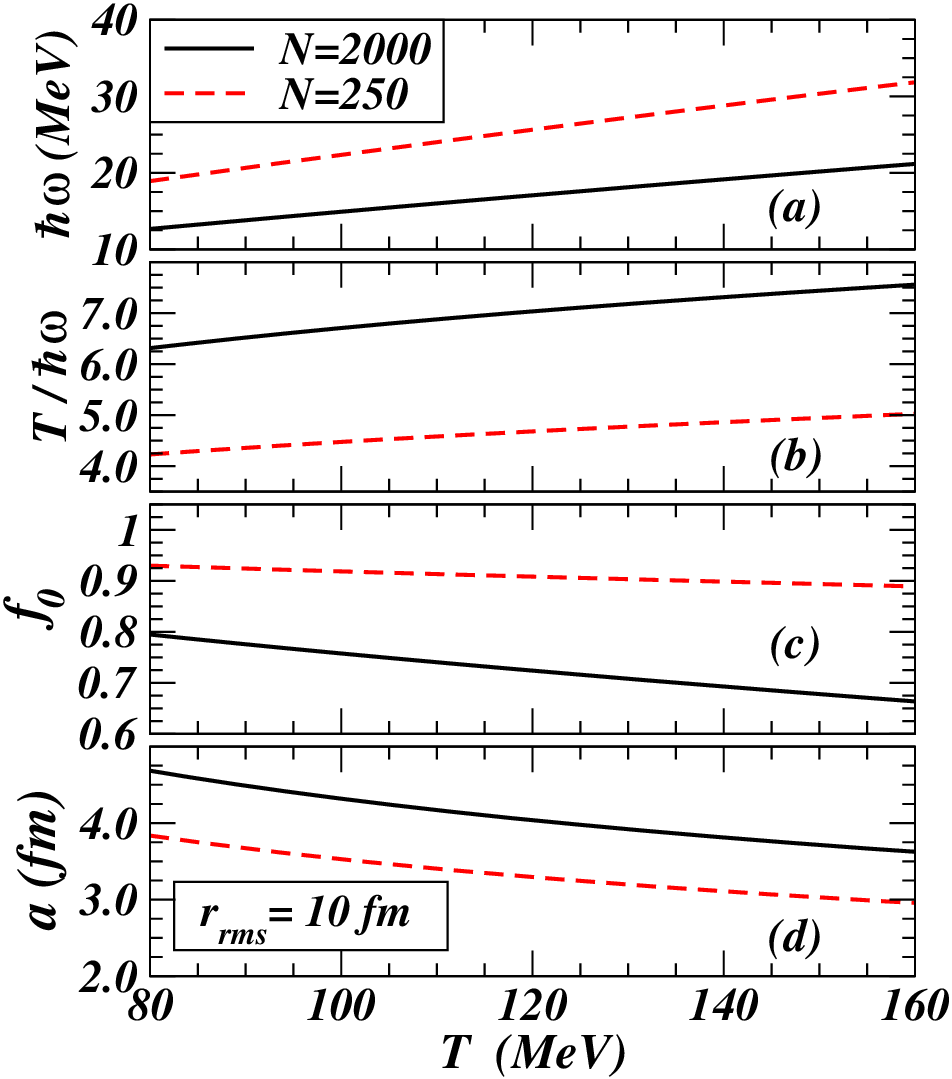}
\caption{(Color online) ($a$) the potential strength $\hbar \omega $,
($b$) the ratio $T/\hbar \omega$, ($c$) the condensate fraction $f_0$,
and ($d$) the oscillator length parameter $a$ for nonrelativistic
boson systems with $N=2000$ and $N=250$ in a static equilibrium with a
$r_{\rm rms}=10$ fm, plotted as a function of temperature.  }
\end{figure}

What is the depth of the mean-field potential that holds the pions
together in static equilibrium for a given $r_{\rm rms}$?  The depth
of the potential is approximately $\hbar \omega (r_{\rm rms}/a)^2 /2$
(see Eq. (\ref{VV})).  For $r_{\rm mrs}=$ 10 fm and $T\sim 120$ MeV,
the results in Figs. 8($a$) and 8($c$) show that the depth of the
potential needs to be about $18 ({\rm MeV})\times 2.5^2/2\sim 56 $ MeV
for $N=2000$, and is about $25 ({\rm MeV})\times 3^2/2\sim 112$ MeV
for $N=250$.  These are not very deep potentials.  It will be of
interest to determine theoretically the mean-field potential for an
assembly of pions at different temperatures.

We reach the following conclusion from the above study: if a
nonrelativistic pion system maintains a static equilibrium within its
mean field, and if it contains a root-mean-squared radius, a pion
number, and a temperature typical of those in high-energy heavy-ion
collisions, then it will contain a large fraction of the Bose-Einstein
pion condensate.  The pion condensation will affect the parameter
$\lambda$ in momentum correlation measurements.

The evolution of pions in high-energy heavy-ion collisions involves
dynamical motion and may not be in a state of static equilibrium.  The
static solutions examined here serve as supplementary tools relative
to which the effects of the dynamical motion and non-equilibrium
effects may be investigated.

\section{Relativistic Treatment of a Boson Gas in a Harmonic Oscillator}

For pions in the environment of a high-energy heavy-ion collision, the
pion temperature is of the order of the pion rest mass and a
relativistic treatment of the pion motion is needed.  We therefore
examine a boson in an external field characterized by a time-like
vector interaction $A_0(r)$, a space-like interaction ${\bf A(r)}$ and
a scalar interaction $S(r)$.  The Klein-Gordon equation for the motion
of the boson is
\begin{eqnarray}
\biggl \{ [p_0-A_0(r)]^2-[{\bf p}-{\bf A}(r)]^2-[m+S(r)]^2 \biggr \}
u(r)=0.
\end{eqnarray}
Different types of interaction potentials will lead to different
single-particle spectra and different Bose-Einstein condensations that
will need to be explored in more detail in the future.  We shall
examine here at this stage only the simplest kind of exactly solvable
potential that is closely connected to the harmonic oscillator
potential in the nonrelativistic limit.  Accordingly, we study scalar
interactions $S(r)$ and introduce the interaction interaction $V(r)$
related to $S(r)$ by
\begin{eqnarray}
V({\bf r})=S({\bf r})+\frac{[S({\bf r})]^2}{2m}.
\end{eqnarray}
The $V({\bf r})$ and the $S({\bf r})$ potentials approach each
other in the nonrelativistic limit of $m\to\infty$. In terms of
$V({\bf r})$, we have
\begin{eqnarray}
[m+S({\bf r})]^2=m^2+2mV({\bf r}),
\end{eqnarray}
and the eigenvalue equation for
relativistic motion with only a scalar interaction becomes
\begin{eqnarray}
\left \{ \frac{{\bf p}^2}{2m}+V({\bf r}) \right \} u ({\bf r})
=\frac{p_0^2-m^2}{2m} u ({\bf r})\equiv \epsilon u ({\bf r}),
\end{eqnarray}
where the eigenvalue $\epsilon$ is related to the particle energy
$p_0$ by
\begin{eqnarray}
p_0 \equiv E = \sqrt{m^2+2m\epsilon}.
\end{eqnarray}
To make the problem simple and to connect with earlier exactly
solvable nonrelativistic solutions, we choose to consider $V(r)$ to
be the same harmonic oscillator potential of Eq.\ ({\ref{VV}),
\begin{eqnarray}
V({\bf r})=\frac{1}{2}m\omega^2 r^2.
\end{eqnarray}
The eigenenergy of the relativistic boson is exactly soluble and is
\begin{eqnarray}
E_n=\sqrt{m^2+2m\epsilon_n} ~,
\end{eqnarray}
where 
\begin{eqnarray}
\epsilon_n=(n+\frac{3}{2}) \hbar \omega.
\end{eqnarray}
We likewise introduce the recalibrated ${\tilde E}_n$ measured
relative to the energy of the $n=0$ state
\begin{eqnarray}
{\tilde E}_n=\sqrt{m^2+2m(n+\frac{3}{2} \hbar \omega)}
-\sqrt{m^2+2m \times \frac{3}{2} \hbar \omega}.
\end{eqnarray}
Instead of the nonrelativistic condition of Eq. (\ref{eq6}), the
relativistic condensate configuration condition becomes
\begin{eqnarray}
\label{releq6}
N=N_0+N_T =\frac{z}{1-z}+\sum_{n>0}^\infty \frac {g_n z e^{- \beta {
\tilde E}_n}} {1-z e^{- {\beta \tilde E}_n}},
\end{eqnarray}
where $g_n=(n+1)(n+2)/2$.

\begin{figure}[h]
\includegraphics[scale=0.5]{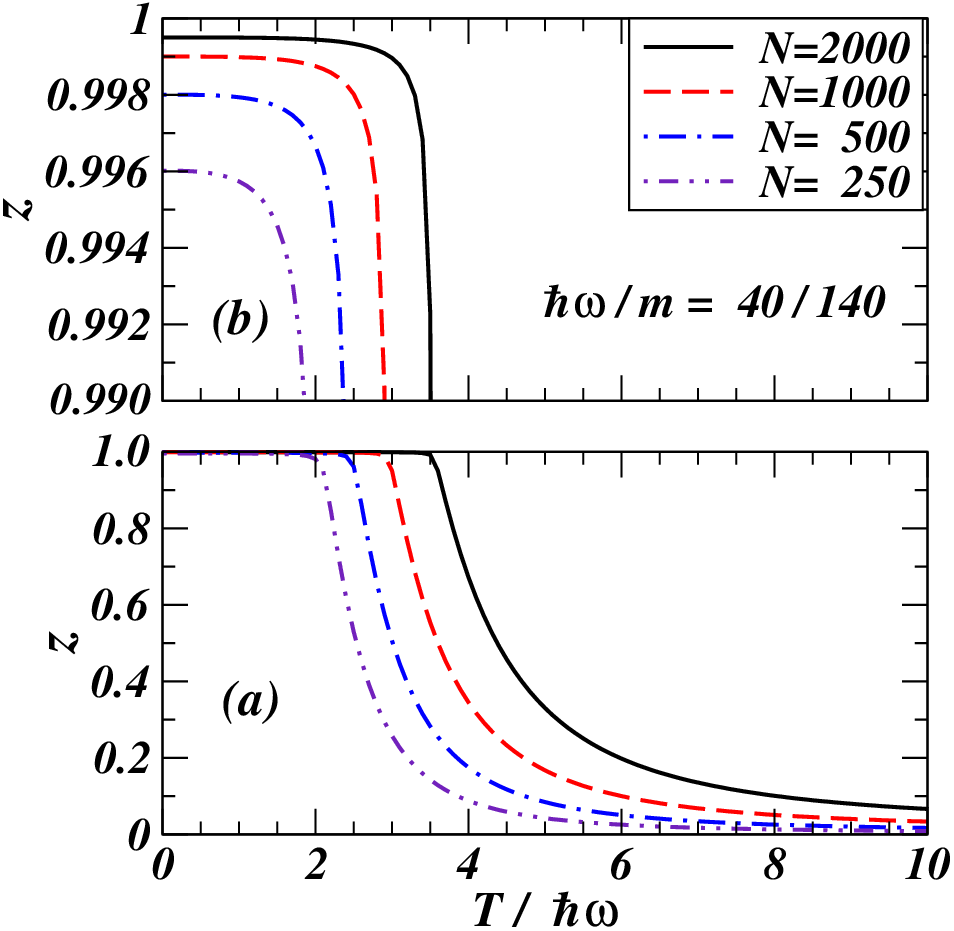}
\caption{(Color online) The fugacity parameter $z$ satisfying the relativistic
condensate configuration condition Eq.\ (\ref{releq6}) for different
boson numbers $N$ and $\hbar \omega/m=20/140$, as a function of the
temperature $T/\hbar \omega$.  Fig. 9($b$) is an expanded view of
Fig.\ 9($a$) in the $z\sim 1$ region.  }
\end{figure}

To solve for the fugacity parameter $z$ in the relativistic case, it
is necessary to specify $\hbar \omega /m$, the ratio of harmonic
oscillator energy scale $\hbar \omega$ to the rest mass $m$ of the
boson.  A small $\hbar \omega/m$ ratio approaching zero corresponds to
the nonrelativistic limit and a large ratio relative to zero leads to
the relativistic case.

We are interested in the case where the mass $m$ of the boson is of
the order of the gas temperature $T$.  We shall see in Fig. 13$a$
below that a boson system with $N=2000$ pions, $T\sim m$, and a
root-mean-squared radius of 10 fm corresponds to a harmonic oscillator
energy $\hbar \omega \approx 40$ MeV which is a substantial fraction of
the rest mass $m$.  We shall therefore investigate relativistic boson
systems with $\hbar \omega/m=40/140$ in our numerical studies.

With this specification of $\hbar\omega/m$ while $N$ and $\beta \hbar
\omega=\hbar\omega /T$ are held fixed, the relativistic condensate
configuration condition (\ref{releq6}) can be solved numerically to
determine the unknown $z$.  We show in Fig.\ 9 the fugacity $z$ which
satisfies the relativistic condensate configuration condition for
different temperatures $T/\hbar \omega$ and different boson numbers
$N$.  To get a better view of the $z$ values, we show an expanded view
of Fig.\ 9($a$) in the $z\sim 1$ region in Fig. 9($b$).

We observe that the fugacity parameter $z$ is close to unity in the
strongly coherent region at low temperatures. Upon a comparison of
Fig.\ 9 with Fig.\ 1, one notices that the shapes of $z$ as a function
of $T/\hbar \omega$ for the relativistic and the nonrelativistic
cases are very similar, except that the scale of the temperatures are
much reduced for the relativistic case.  For $N=2000$, the condensate
temperature occurs at $T/\hbar \omega \sim 3.5$ in the relativistic
case, in contrast to the nonrelativistic case at $T/\hbar \omega \sim
11$.  For $N=250$, the condensate temperature occurs at $T/\hbar
\omega \sim 1.9$ in the relativistic case, in contrast to the
nonrelativistic case of $T/\hbar \omega \sim 5.1$.  To see why these
large changes occur, we note that the recalibrated energy expanded in
powers of $1/m$ is
\begin{eqnarray}
{\tilde E}_n={\tilde \epsilon}_n -
\frac{[(n+3/2)\hbar \omega]^2}{2m}+O(\frac{\epsilon_n^3}{m^2}).
\end{eqnarray}
For the relativistic harmonic oscillator potential we have chosen, the
spectrum of ${\tilde E}_n$ is nearly the same as those in the
nonrelativistic case of ${\tilde \epsilon}_n$ for small values of $n$.
However, the spectrum for large values of $n$ is greatly compressed by
the presence of the second term with a negative sign in the above
equation.  As a result, a large number of chaotic particles can be
accommodated even at a lower temperature, leading to a large shift of
the condensate temperature in units of $\hbar \omega$ in Fig.\ 9 when
relativistic effects are included.

\begin{figure}[h]
\includegraphics[scale=0.5]{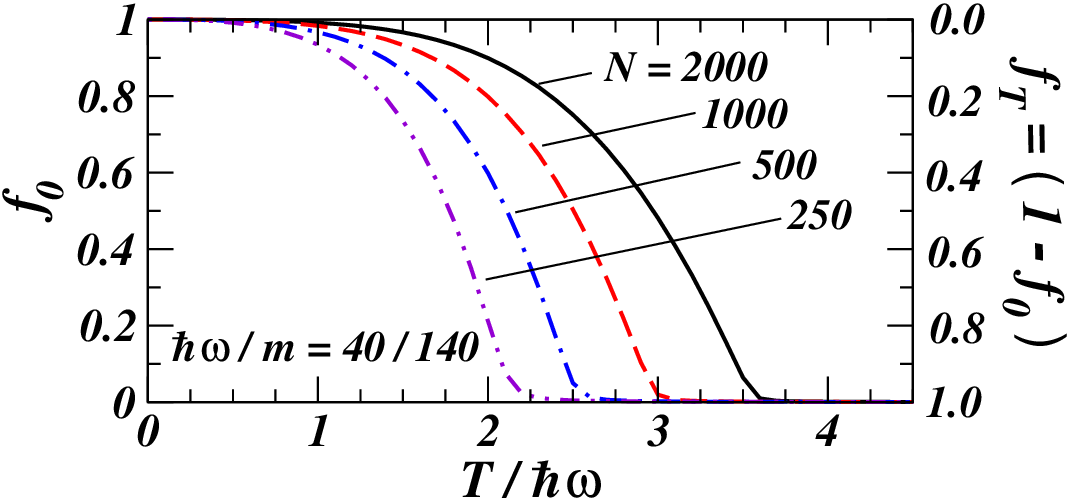}
\caption{(Color online) Different curves represent condensate
fractions $f_0(T)$ as a function of $T/\hbar \omega$ for different
boson numbers $N$, calculated with the relativistic condensate
configuration condition Eq.\ (\ref{releq6}) for
$\hbar\omega/m=40/140$.  The abscissa labels for the corresponding
chaotic fraction $f_T(T)=[1-f_0(T)]$ are indicated on the right.  }
\end{figure}

After the value of the solution $z$ is obtained, $N_0=z/(1-z)$ and
$N_T$ can be subsequently determined to give the condensate
configuration specified by the condensate fraction $f_0$ and the
chaotic fraction $f_T$.

The condensate fractions $f_0(T)$ as a function of $T/\hbar \omega$
calculated with the fugacity parameters of Fig.\ 9 for different boson
numbers $N$ and $\hbar\omega/m=40/140$ are shown as different curves
in Fig.\ 10.  The abscissa labels for the corresponding chaotic
fraction $f_T(T)=[1-f_0(T)]$ are indicated on the right. We observe
that the behavior of the condensate fraction in the relativistic case
is similar to the nonrelativistic case, with the exception of the
shift of temperature $T/\hbar\omega$ to lower values.  Again, the
transition from the condensate phase to the chaotic phase occurs over
a large range of temperatures and is therefore not a sharp first-order
type transition.  The complementary chaotic fraction $f_T(T)$
increases gradually as the temperature increases, reaching the value
of unity at large $T/\hbar \omega$.

\section{Spatial and momentum distributions in the Relativistic case}

\begin{figure}[h]
\includegraphics[scale=0.5]{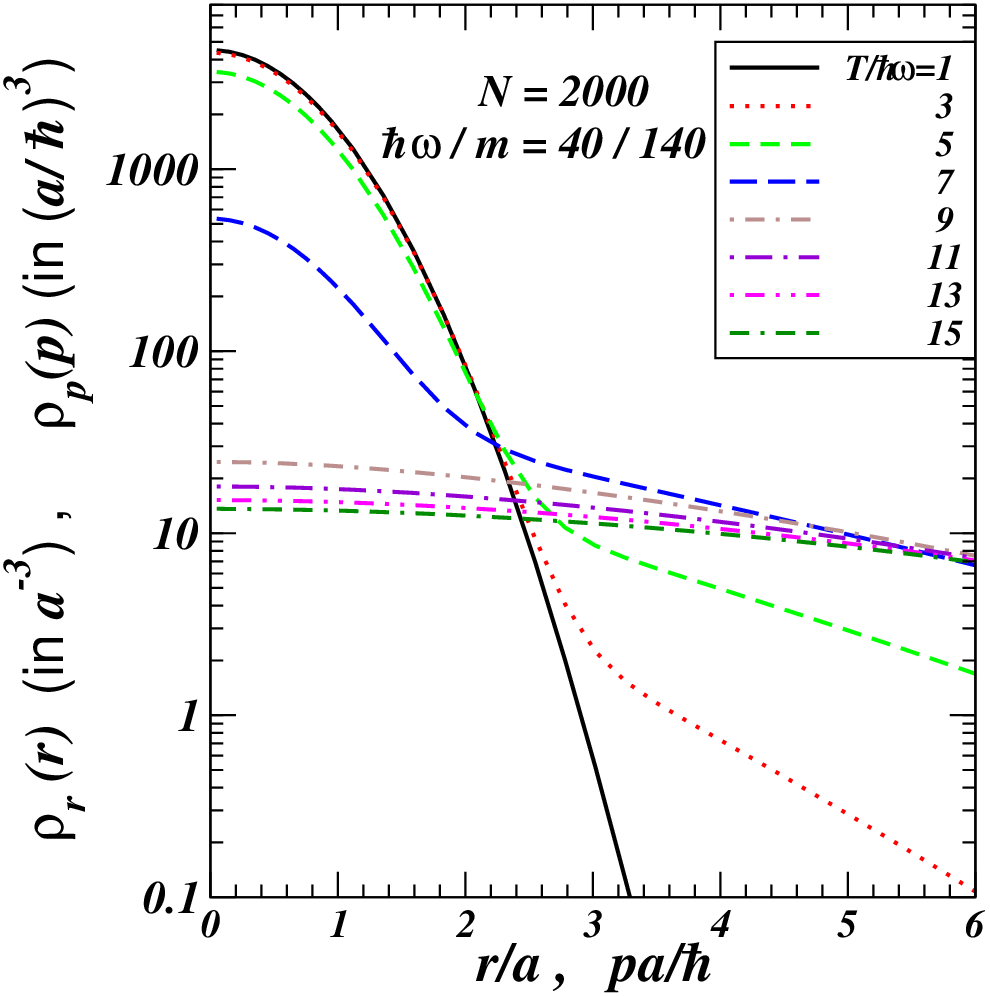}
\caption{(Color online) The spatial density distribution
$\rho_r(\bbox{r})$ in units of $a^{-3}$, expressed as a function of
$r/a$, and the momentum density distribution $\rho_p(\bbox{p})$ in
units of $(a/\hbar)^3$, expressed as a function of $pa/\hbar$, for the
relativistic case of $\hbar\omega/m=40/140$ with $N=2000$ at different
temperatures.  }
\end{figure}

The knowledge of the fugacity parameter for different temperatures
allows one to determine the occupation numbers at different
single-particle states.  These occupation numbers and the absolute
square of the single-particles wave functions give the spatial and
momentum densities of the system at different temperatures.  As we
remarked previously, $\rho_p(\bbox{p})$ and $\rho_r(\bbox{r})$ have
the same shape when properly scaled. It suffices to consider the
spatial density $\rho_r(\bbox{r})$ given by
\begin{eqnarray}
\rho_r(\bbox{r})=G^{(1)}(\bbox{r},\bbox{r})
=\sum_{n=0}^\infty \frac { z e^{- \beta {
\tilde E}_n}} {1-z e^{- {\beta \tilde E}_n}}
u_n^*({\bf r}) u_n({\bf r}),
\end{eqnarray}
where $n$ represents the set of quantum numbers $\{ n_r l m \}$ of a
harmonic oscillator state, $u_n({\bf r})$ is the harmonic oscillator
wave function normalized to $\int d{\bf r} |u_n({\bf r})|^2=1$,
\begin{eqnarray}
u_n({\bf r})=N_{n_r l} x^l e^{-x^2/2}
L_{n_r}^{l+\frac{1}{2}}(x^2) Y_{lm}(\theta,\phi),
\end{eqnarray}
 $n=2n_r+l$, $x=r/a$, $L_{n_r}^{l+\frac{1}{2}}(x^2)$ is the associated
Laguerre polynomial, and
\begin{eqnarray}
(N_{n_r l})^2=\frac{2 n !}{a^3 \Gamma(n_r+l+\frac{3}{2})}.
\end{eqnarray}

\begin{figure}[h]
\includegraphics[scale=0.5]{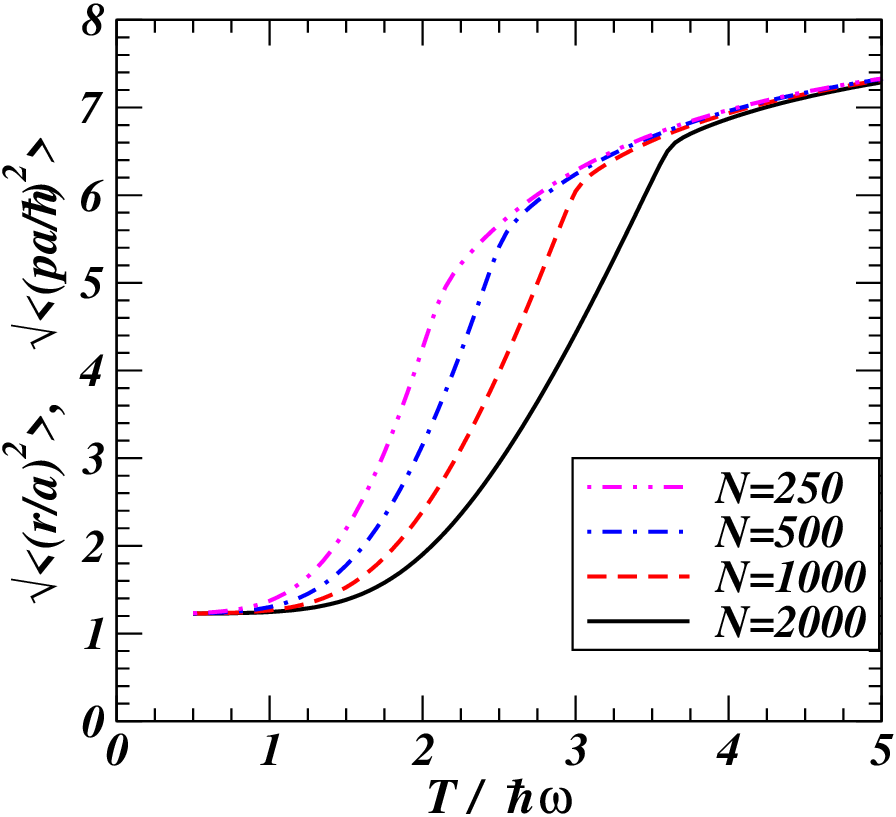}
\caption{(Color online) The root-mean-squared radius in unit of $a$
and the root-mean-squared momentum in units of $\hbar/a$, as a
function of $T/\hbar\omega$ for different numbers of bosons in the
system.}
\end{figure}

We plot in Fig. 11 the spatial and momentum distributions of the
system with $N=2000$ for the relativistic case of
$\hbar\omega/m=40/140$ as a function of their dimensionless variables
${r}/a$ and ${p}a/\hbar$ respectively.  One observes that up to
$T/\hbar\omega\sim 3 $ the system has a small spatial or momentum size
and there is a substantial condensate fraction in the system.  In
Fig. 12 we plot the root-mean-squared radius in unit of $a$, $r_{\rm
rms}/a=\sqrt{\langle (r/a)^2\rangle}$, and the root-mean-squared
momentum in unit of $\hbar/a$, $p_{\rm rms}a/\hbar=\sqrt{\langle
(pa/\hbar)^2\rangle}$, as a function of $T/\hbar \omega$.  For
$N=2000$, the quantity $r_{\rm rms}/a$ is slightly greater than 1 up
to $T/\hbar \omega \sim 2$, and it increases relatively rapidly to
about 6.5 at the condensate temperature, $T_c/\hbar \omega \sim 3.5$.
It increases at a relatively slower rate at temperatures above $T_c$.

\begin{figure}[h]
\includegraphics[scale=0.5]{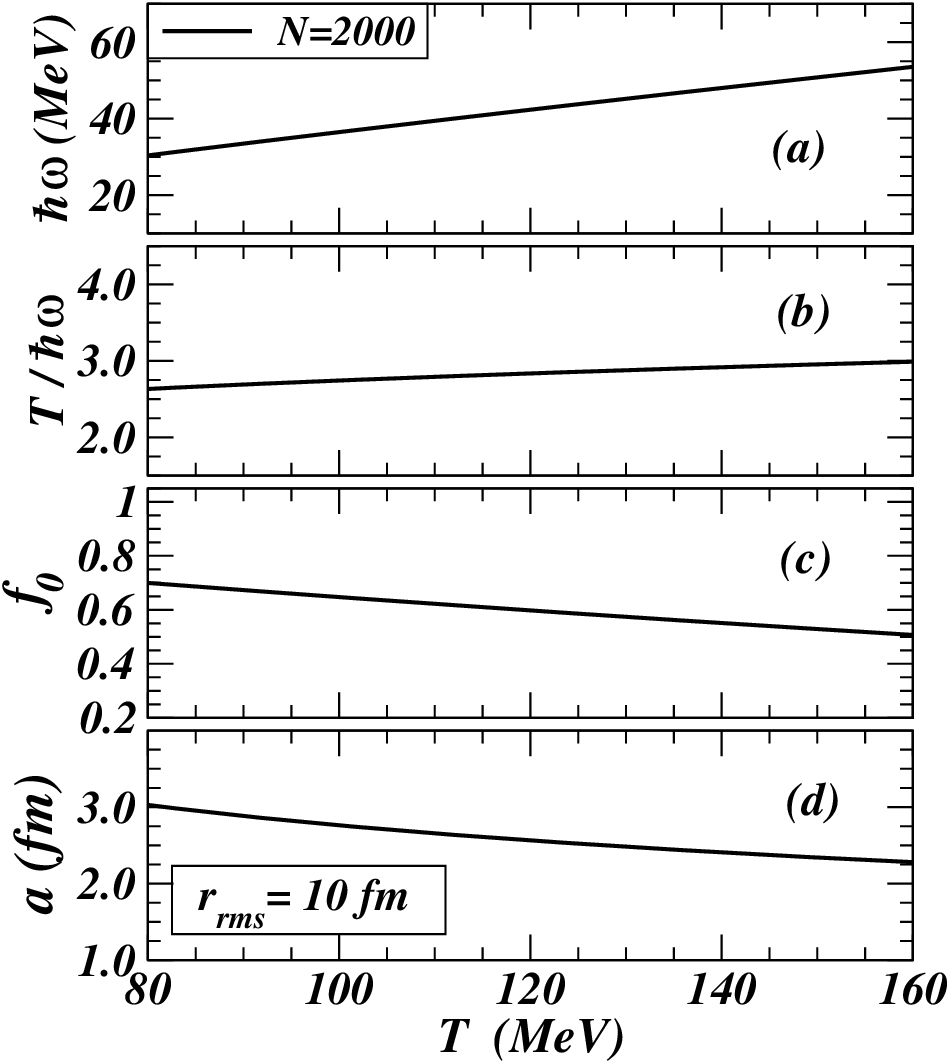}
\caption{ ($a$) the potential strength $\hbar \omega $, ($b$) the
ratio $T/\hbar \omega$, ($c$) the condensate fraction $f_0$, and ($d$)
the oscillator length parameter $a$ for relativistic boson systems
with $N=2000$ and $N=250$ in a static equilibrium with a $r_{\rm
rms}=10$ fm, plotted as a function of temperature.  }
\end{figure}

We can carry out an analysis to inquire the following: If a system of
$N=2000$ relativistic pions is held together by its mean field, taken
to be a harmonic oscillator, and if it comes to a state of static
equilibrium with a given root-mean-squared radius $r_{\rm rms}$ at a
temperature $T$, what is the condensate fraction of such a system?  We
shall therefore examine a pion system with $r_{\rm rms}=10$ fm, a
temperature range from 80 to 160 MeV, and the number of pions to be
$N=2000$ (for a central RHIC Au-Au collisions at $\sqrt{s_{NN}}=200$
GeV).  We have chosen $\hbar \omega/m=40/140$, to be approximately
self-consistent with the value of $\hbar \omega$ extracted form such
an analysis (see Fig. 13($a$)).

If the value of $r_{\rm rms}$ is fixed as given, the quantities $\hbar
\omega $ and $T$ are then related by the set of parametric equations
of (\ref{pareq1}) and (\ref{pareq2}).  Using the function $r_{\rm
rms}/a=F_N(x)$ of Fig. 12 in the above equations and varying
$x=T/\hbar \omega$ for a fixed $r_{\rm rms}$, one can determine the
energy $\hbar \omega $ as a function of $T$ for the case of $N=2000$.
The results are shown in Fig. 13$(a)$.  One finds that for the pion
system with a given root-mean-squared radius of 10 fm, the value of
$\hbar \omega $ ranges from about 30 to 53 MeV for $N=2000$, with an
average of about 42 MeV.  The ratio of $T/\hbar\omega $ is about 2.7
to 3 for $N=2000$ as shown in Fig. 8($b$).  From these ratios of
$T/\hbar\omega$, one can use Fig. 2 to find out the condensate
fraction.  The condensate fractions $f_0(T)$ for a pion gas at various
temperatures with $N=2000$ are shown in Fig. 8($c$).  One finds that
$f_0(T)$ is between 0.5 and 0.7 for $N=2000$.  The knowledge of $\hbar
\omega$ in Fig. 8($a$) allows one to determine the values of $a$ as a
function of temperature as shown in Fig. 8($d$).  The oscillator
length $a$ is between 3 and 2.2 fm for $N=2000$.

The relativistic analysis indicates that the relativistic effects
change the single-particle spectrum and shift the locations of the
condensate fraction in units of $\hbar \omega$.  The condition of
maintaining a system size with a root-mean-squared radius of 10 fm
recalibrates and raises the oscillator energy $\hbar \omega$ for the
relativistic case, as compared to the nonrelativistic case.  As a
consequence, the condensate fraction for $N=2000$ is modified from
$f_0\sim 0.67$ to 0.8 in the nonrelativistic to $f_0\sim0.5$ to 0.7
in the relativistic case.  There is a small reduction of the
condensate fraction, but the condensate fraction remains quite large
in the relativistic case.
 
We again reach the following conclusion from this study: If a
relativistic pion system maintains a static equilibrium within its
mean field, and if it contains a root-mean-squared radius, a pion
number, and a temperature typical of those in high-energy heavy-ion
collisions at RHIC, then it will contain a large fraction of the
Bose-Einstein pion condensate.  

\section{Discussions and Conclusions} 

As the chaoticity parameter $\lambda$ has been widely used in all HBT
measurements, we are therefore motivated to investigate an exactly
solvable problem to study the momentum correlation function for a
noninteracting boson gas assembly held together in a harmonic
oscillator potential at various temperatures.  In the process, we find
that the phase transition from the Boson-Einstein condensate to the
chaotic phase occurs gradually over a large range of temperatures,
with the condensate fraction $f_0(T)$ varying approximately as
$1-(T/T_c)^3$, where the condensate temperature $T_c$ is approximately
given by $(N/1.202)^{1/3} \hbar \omega$.  The spatial and the momentum
radii of the system are small in a condensate at low temperatures, of
the order of a few oscillator units, increasing in size as the
temperature reaches the chaoticity limit.

From the momentum correlation function, we can determine the
$\lambda(p,T)$ parameter and the HBT radius $R_{\rm HBT}$.  We find
that the $\lambda(p,T)$ parameter is a sensitive function of both the
pair momentum $p$ and temperature $T$.  For a temperature above the
condensate temperature, the $\lambda(p,T)$ parameter is 1 for all
momentum $p$.  However, for temperatures below and even substantially
below the condensate temperature, $\lambda(p,T)$ is small and close to
zero for small pair momentum $p$, but it increases and saturates at
$\lambda(p,T)=1$ at large pair momentum $p$.  The location where
$\lambda(p,T)$ attains unity changes with temperature.  The lower the
temperature, the greater is the value of $p$ at which the
$\lambda(p,T)$ attains the value of unity. Because the $\lambda(p,T)$
parameter attains the value of unity for systems at temperatures much
below the condensate temperature, the occurrence of $\lambda=1$ is not
consistently correlated with the absence of a condensate fraction.
Only in the region of small $p$ will the parameter $\lambda(p,T)$ be
correlated with, but not equal to, the chaotic fraction $f_T(T)$ of
the system.

We find that the HBT radius $R_{\rm HBT}$ increases gradually with
increasing pair momentum $p$ and temperature $T$.  However, for small
value of $p$, the HBT radius increases only slowly with increasing
temperature at low temperatures and it then increases rapidly and
abruptly as the temperature approaches the condensate temperature.
The temperature dependence of the HBT radius at small $p$ values
correlates well with the temperature dependence of the
root-mean-squared radius of the system.

It is of interest to inquire the degree of coherence of pion systems
produced in high-energy heavy-ion collisions.  We have examined both
cases of pions as a nonrelativistic and a relativistic gas in a
harmonic oscillator potential.  If a pion system maintains a static
equilibrium in its mean field, and if it contains pion numbers from
$N=250$ to $N=2000$, a temperature in the range from 80 to 160 MeV,
and a root-mean-squared radius of 10 fm, (typical of those one
encounters in high-energy heavy-ion collisions), then it will contain
a large fraction of the Bose-Einstein pion condensate.  While the
details of the nonrelativistic and relativistic calculations are
presented in Sections VIII and X, we can provide simple arguments here
to indicate that these are reasonable results based on plausible
physical principles.  Bose-Einstein condensation occurs when the
temperature is below the condensate temperature, which is a few units
of $\hbar \omega$.  We need to estimate the energy scale $\hbar
\omega$ for the pion system.  The energy scale can be estimated by
knowing the length unit $a$.  One expects that the pion system with a
root-mean-squared radius of 10 fm would be contained within a few
units of this length $a$, leading to a rough estimate of the length
unit $a$ to be about a few fm.  By dimensional analysis, the energy
scale associated with this length unit $a$ for a pion is $\hbar
\omega=\hbar^2/ma^2$, which gives a value many tens of MeV for
$\hbar\omega$.  With a temperature of $T=120$ MeV or 140 MeV, we
obtain the ratio $T/\hbar \omega$ of a few units which would
correspond to a $T/\hbar \omega$ ratios with a substantial condensate
fraction.  We can therefore understand that the occurrence of the pion
condensation in static equilibrium arises because the pions are
massive particles, and a large number of pions are produced and
concentrated in a small spatial volume characterized by a
root-mean-squared radius of only 10 fm.  The pion gas in static
equilibrium is therefore in the realm of low-temperature boson systems
with possible occurrence of Bose-Einstein condensation.

The evolution of pions in high-energy heavy-ion collisions contain
dynamical motion and may not be in a state of static equilibrium.  How
the dynamical motion of the pions will modify the coherence of the
system will be an interesting subject for future investigations.  

While we await future theoretical investigations, it is of interest in
the meantime to discuss possible modifications of the static results
obtained here in the presence of a collective expansion.  One expects
that the collective expansion will not alter the energy ordering of
the states of the system.  Because the Bose-Einstein condensation
depends on the relative ordering of the energy of the states, the
coherence may not be greatly affected.  However, the average pair
momentum in the direction of the expansion will be greatly boosted.
The $\lambda(p,T)$ parameter for the expanding coherent source as a
function of the pair momentum would likely retain a shape similar to
the static case, but with the $p$ boosted by the collective expansion.
In this connection, it is interesting to note that the experimental
$\lambda$ values plotted as a function of the pair transverse momentum
has a shape \cite{Ada05,Adl04} quite similar to the shape of the
$\lambda(p,T)$ parameter plotted as a function of the average pair
momentum $pa/\hbar$ in Fig.\ 6($a$).  Although alternative
explanations in terms of a decrease in the resonance decay
contributions at higher $p_T$ have been presented, it will be of
interest to explore whether the behavior of $\lambda$ as a function of
the pair transverse momentum may be due to the occurrence of an
expanding Bose-Einstein condensate.

The theoretical HBT radius for a static source increases slightly as
the pair momentum increases, while the experimental measurements gives
an HBT radius decreasing as $p_T$ increases.  The theoretical HBT
radius may be more sensitively affected by the expansion dynamics
because the collective expansion boosts not only the average pair
momentum but also the relative momentum between the correlated pair,
the boost being the greater the larger the magnitude of the pion
momentum.  As a larger relative momentum leads to a smaller HBT
radius, the HBT radius therefore decreases as a function of the pair
momentum.  Clearly, whether future analyses bear out this possibility
will be of great interest.  How the collective pion motion will affect
quantitatively the HBT radius for a boson system with varying degrees
of coherence is therefore an interesting subject for future
investigations.

It has been proposed that the question of whether an observation of
$\lambda < 1$ is due to coherence or due to contamination from
particles from far outside the source volume can be tested by
analyzing three-particle correlations \cite{Hei97}.  Such analyses of
data at both SPS and RHIC have been consistent with the chaotic
conjecture \cite{Bog99}.  However, as we note that the $\lambda(p,T)$
parameter can assume the value of unity in certain kinematic regions
even for significantly coherent systems with a temperature much below
the condensate temperature, the attainment of $\lambda=1$ cannot be a
unique signature of the chaoticity of a system.  However, how the
coherence of the boson system may affect three-body correlations has
not been worked out explicitly and merits further investigations to
clarify the situation.

     It needs to be emphasized that  to make the problem
     tractable as an exactly solvable model, we have specialized to a
     static treatment of the boson system in equilibrium in both a
     nonrelativistic and a relativistic harmonic oscillator
     potential, for which analytical eigen-energies and eigenfunctions
     can be readily available.  This is a simple model of
     noninteracting boson gas in an external potential without
     two-body interactions.  Even with such an idealization, a wealth
     of new information on the coherence and two-particle momentum
     correlation functions as well as the chaoticity parameter has
     been obtained as a function of the attributes of the boson
     environment.

     The harmonic oscillator potential introduced here can arise from
     an external trap, as in atomic physics.  In high-energy heavy-ion
     collisions, the harmonic oscillator potential can arise
     approximately from the mean-field potential experienced by a
     pion, due to the interactions generated by other pions and medium
     particles.  While approximating the pion mean-field potential as
     a harmonic oscillator potential can yield gross features and a
     wealth of information, a more accurate determination of the pion
     momentum and correlation functions will require a better
     description of the pion mean-field potential. As the mean-field
     potential depends on the pion density as in the Glauber theory
     \cite{Gla59}, and the equilibrium pion density depends in turn on
     the mean-field potential, it will be necessary in the future to
     study the pion mean-field potential and the density
     self-consistently in a pion condensate. Besides these mean-field
     interactions between pions, the remaining residue interactions
     will give rise to additional complications which may be studied
     in the future.

     We have described the correlations in a static equilibrium
     environment and we need to discuss how the time dependence can be
     handled.  An accurate dynamical treatment will examine the time
     evolution of the system for a set of given initial conditions, as
     in a time-dependent Hartree approximation following the technique
     of the time-dependent Hartree-Fock approximation developed in
     nuclear physics \cite{Bon76,Won75}

     While we await future work on the time dependence of the
     correlation function, the static results obtained here can be
     used for experimental comparison if the time dependence of the
     external field is such that the external potential is suddenly
     removed, as in a typical condensed matter experiment with trapped
     atomic particles in a condensate.  In these low-temperature
     measurements with atoms, the trapped atoms before being released
     are now described as having an equilibrium momentum distribution
     in momentum space, appropriate for the system in a given
     external field at a given temperature.  The sudden removal the
     external field allows the initial momentum distribution of the
     particle to be frozen at the moment of the external field
     removal, as appropriate under the application of the sudden
     approximation in quantum mechanics.  Subsequent free streaming of
     the particles without the external field and mutual interactions
     (except for the additional correction of the gravitational field
     or other extra forces applied to the particles) allows the
     reconstruction of the momentum distribution of the source at the
     moment of its freezing out.  In measuring the arrival times and
     arrival positions of the particles of a correlated pair in
     Ref. \cite{Sch05}, the quantities that are in effect measured are
     the momenta of correlated pairs from which the average momenta
     and the relative momenta of the pair can be collected and
     examined.  The perspectives of studying the correlation in
     momentum space presented here offer useful complementary
     viewpoints to the theoretical and experimental works in atomic
     physics have been focused so far on the correlation function in
     configuration space.

     The results obtained here can be approximately applied to
heavy-on collisions if the explosive expansion is so rapid that it can
be approximately described as a sudden removal of the external field.
In that case, the static initial momentum distribution and
correlations of the particles would be frozen at the moment of the
external field removal and show up as particles reaching the detectors
by free streaming.  In this respect, it is of great interest to
examine in the future a dynamical model of the expansion of the pion
gas and study how the explosive expansion will affect the momentum
correlation function.

\begin{acknowledgments}
The authors would like to thank Prof.\ R.\ Glauber for stimulating
discussions and for pointing out the importance of the pion coherence
in high-energy heavy-ion collisions.  The authors wish to thank Drs.\
Teck-Ghee Lee and Jian-Shi Wu for helpful discussions.  This research
was supported in part by the National Science Foundation of China
under Contract No. 10575024, and in part by the Division of Nuclear
Physics, Department of Energy, under Contract No. DE-AC05-00OR22725
managed by UT-Battelle, LLC.
\end{acknowledgments}

\end{document}